\documentclass[12pt,letterpaper]{article}
\usepackage[usenames, dvipsnames]{xcolor}
\usepackage{jcapmod}

\usepackage{tocloft}
\usepackage[]{todonotes}
\usepackage{verbatim}
\usepackage{amsmath}
\usepackage{mathrsfs}
\usepackage{cleveref}

\usepackage{amsthm}
\usepackage{tikz}

\usepackage{tikz}
\usepackage{setspace,caption}
\usepackage{lipsum}
\usetikzlibrary{matrix,arrows,calc}

\usepackage{bbm} 					
\usepackage{slashed} 				
\usepackage{graphicx}				
\usepackage{subcaption}			
\usepackage{psfrag}				
\usepackage{tensor}				
\usepackage{fouridx}				
\usepackage{bm}					
\usepackage{mdframed}				
\usepackage{multirow}				
\usepackage{soul}					
\usepackage{bbold}				
\usepackage{multicol}				
\usepackage{tikz-cd}
\usepackage{rotating}

\usetikzlibrary{arrows}

\tikzset{
commutative diagrams/.cd,
arrow style=tikz,
diagrams={>=latex}}

\usepackage{amsmath}
\usepackage{amssymb}
\usepackage{mathtools}

\usepackage{feynmf}
\usepackage{marvosym}

\usepackage{import}

\usepackage[dvipsnames]{xcolor}
\usepackage[utf8]{inputenc}
\usepackage{amsmath}
\usepackage{tikz}
\usepackage{extpfeil}
\usepackage{amsfonts}
\usepackage{amssymb} 
\usepackage{bm}
\usepackage{graphicx}
\usepackage{array}
\usepackage{lipsum}
\usepackage{float}
\usepackage{multirow}
\usepackage{hhline}
\usepackage{tabularx}

\usepackage{graphicx}
\usepackage{afterpage}
\usepackage{longtable}
\usepackage{epstopdf}
\usepackage{amsthm}

\newcommand{\bZ}{\mathbb{Z}}

\def\ge{E}
\def\gso{SO}
\def\gsu{SU}
\def\gsp{Sp}
\def\gf{F}
\def\gg{G}

\newcommand{\XXX}[3]{}

\newcommand*\xoverline[2][0.75]{%
    \sbox{\myboxA}{$\m@th#2$}%
    \setbox\myboxB\null
    \ht\myboxB=\ht\myboxA%
    \dp\myboxB=\dp\myboxA%
    \wd\myboxB=#1\wd\myboxA
    \sbox\myboxB{$\m@th\overline{\copy\myboxB}$}
    \setlength\mylenA{\the\wd\myboxA}
    \addtolength\mylenA{-\the\wd\myboxB}%
    \ifdim\wd\myboxB<\wd\myboxA%
       \rlap{\hskip 0.5\mylenA\usebox\myboxB}{\usebox\myboxA}%
    \else
        \hskip -0.5\mylenA\rlap{\usebox\myboxA}{\hskip 0.5\mylenA\usebox\myboxB}%
    \fi}
\makeatother

\newcommand{\email}[1]{\href{mailto:#1}{#1}}

\begin{document}
	\newcommand{\main}{.}
\begin{titlepage}

\setcounter{page}{1} \baselineskip=15.5pt \thispagestyle{empty}
\setcounter{tocdepth}{2}

\bigskip\

\vspace{1cm}
\begin{center}
{\fontsize{22}{28} \bfseries On the Scarcity of Weak Coupling \\ \vspace{0.25cm} in the String Landscape }
\end{center}

\vspace{0.45cm}

\begin{center}
\scalebox{0.95}[0.95]{{\fontsize{14}{30}\selectfont James Halverson, Cody Long, and Benjamin Sung}} \vspace{0.25cm}
\end{center}

\begin{center}

\textsl{Department of Physics, Northeastern University, Boston, MA 02115, USA}\\

\vspace{0.25cm}

\vskip .3cm

\email{\tt j.halverson@northeastern.edu, co.long@northeastern.edu, \\ b.sung@northeastern.edu}
\end{center}

\vspace{0.6cm}
\noindent
We study the geometric requirements on a threefold base for the corresponding
F-theory compactification to admit a weakly-coupled type IIB limit. We examine both the
standard Sen limit and a more restrictive limit, and determine
conditions sufficient for their non-existence for both toric bases and
more general algebraic bases. In a large ensemble of geometries
generated by base changing resolutions we derive an upper bound on the
frequency with which a weak-coupling limit may occur, and find that such limits
are extremely rare. Our results sharply quantify the widely held notion
that the vast number of weakly-coupled IIB vacua is but a tiny
fraction of the landscape.

\noindent
\vspace{2.1cm}

\noindent\today

\end{titlepage}
\tableofcontents

\newpage
\section{Introduction}
The landscape of string theory vacua has provided a rich ensemble of
four dimensional effective theories of quantum gravity, and many
candidates of an ultraviolet completion of our
universe~\cite{Bousso:2000xa,Ashok:2003gk,Denef:2004ze}. A large
number of 4d $\mathcal{N}=1$ theories are realized by weakly-coupled
string theory on Calabi-Yau manifolds and orientifolds thereof. Such
compactifications have provided a vast set of four-dimensional
effective theories enumerated by choices of geometry and flux, and
have also yielded many non-trivial computations of operators in the
low energy theory, including various candidates for BSM physics, such
as interesting cosmologies and particle spectra (see e.g.~\cite{MayorgaPena:2017eda} for recent work).  However, it is
believed that this weakly-coupled landscape is only a tiny fraction of
the landscape of 4d theories coupled to gravity that have UV
completions.

In this work we endeavor to determine the prevalence or scarcity of
one of the best studied corners of the landscape, weakly coupled type
IIB orientifold compactifications, within its natural larger context,
F-theory.  F-theory~\cite{Vafa:1996xn,Morrison:1996pp} is an
intrinsically strongly-coupled generalization of type IIB string
theory on a Calabi-Yau orientifold that incorporates a holomorphically
varying axio-dilaton in the structure of an elliptically fibered
Calabi-Yau variety $X$\footnote{We will be interested in $d=4$
  compactifications, and therefore $X$ will be a fourfold.}.  In
some cases a particular limit in the complex structure moduli of $X$,
known as the Sen limit~\cite{Sen:1996vd, Sen:1997bp}, recovers weakly
coupled type IIB on the base $B$ of $X$, where $B$ is realized as an
orientifold of a Calabi-Yau threefold. Interestingly, unless a strict
Sen limit is taken in a way that will be made precise, there are
always regions in $B$ where $g_s$ is $O(1)$, which motivates us to
also study a global weak coupling limit (GWCL) that can avoid the
existence of strongly coupled regions. Whether an F-theory geometry
$X$ admits a limit in complex structure that gives rise to a Sen limit
or GWCL is determined by the geometry of $B$. We study conditions
under which the topology of $B$ forbids such limits, and the frequency
with which these conditions hold in large ensembles of geometries.

Our main result is that there are relatively simple conditions on $B$ that determine whether there exists
a Sen limit or a GWCL, and in the ensembles of bases $B$ that we study the probability that such
a limit exists is extremely small. Specifically, in our ensemble the fraction of bases admitting a GWCL
or a Sen limit are bounded by
\begin{equation}
  \frac{N_{\text{GWCL}}}{N_{\text{Total}}} \leq 1.1 \times 10^{-723}\ \qquad \qquad 
\frac{N_{\text{Sen}}}{N_{\text{Total}}}  \leq  3.0  \times 10^{-391}\,.
\end{equation}
In the discussion at the end we will argue that this probability can only shrink in ensembles
of toric bases $B$ larger than ours, and that the low probability of the existence of these
limits should also be a feature of other ensembles.

These results suggest that the widely studied weakly coupled type IIB
landscape is but a small fraction of F-theory, and
therefore also of the landscape as whole. While its relative scarcity
has been widely anticipated, this attempt to quantify it exacerbates the need
to better understand moduli stabilization in strongly coupled limits
of string theory, as well as
transitions between vacua arising in different low energy effective theories
that themselves arise from different strongly coupled compactifications.

\vspace{.7cm}

At a technical level, our results arise from understanding how
seven-brane configurations are modified under topological
transitions. For example, specific topological transitions from
$B$ to  $B'$ induce a transition of the associated Calabi-Yau fourfold
$X$ to $X'$ that gives rise to so-called non-Higgsable clusters (NHC)
\cite{Morrison:1996pp,Morrison:2012np,Morrison:2012js} with high
probability. These are clusters of non-trivial intersecting
seven-brane configurations that exist for generic values of the
complex structure of $X'$. Moduli stabilization at generic
points in moduli space can therefore achieve gauge symmetry
\cite{Grassi:2014zxa} and there is increasing
evidence \cite{Halverson:2015jua,Taylor:2015ppa,Halverson:2017ffz} that generic bases give rise to such clusters for generic
moduli. For example, in the ensemble of
$\frac43 \times 2.96 \times 10^{755}$ geometries from our previous
work \cite{Halverson:2017ffz}, the fraction of geometries that
exhibited non-Higgsable clusters was
\begin{equation}
\frac{N_\text{NHC}}{N_\text{Total}} \geq 1 - 1.01 \times 10^{-755},
\end{equation}
i.e. essentially every geometry exhibited NHCs.
Furthermore, the list of individual non-Higgsable seven-brane (NH7) structures
that can exist in a cluster is rather short, though the building blocks can be assembled
into a cluster in many ways, and most of the individual seven-brane structures are
intrinsically strongly coupled; see \cite{Halverson:2016vwx} for a detailed study. Together, these facts provide strong evidence
that generic points in the complex structure moduli space of
generic F-theory geometries are strongly coupled.

However, this does not preclude the existence of a Sen limit or a GWCL, since if the gauge factors
on the seven-brane are sufficiently low rank then there could exist limits in moduli space in
which the gauge factor is actually enhanced, and can be described by $N$ D7-branes on top of an O7-plane
at weak coupling. These arise from so-called Kodaira $I_n^*$ singular fibers. This potential loophole is the technical focus of this work, and what we show is that if the low rank strongly coupled NH7 is tuned
to $I_n^*$, those or other seven-branes in the geometry are forced to enhance beyond $I_n^*$, under
simple conditions that occur with high probability.  These more singular seven-branes carry exceptional
gauge groups and forbid either of the weak coupling limits.

The outline of this paper is as follows. In \S\ref{sec:weakcup} we review the standard Sen limit, and
introduce a global weak coupling limit that ensures $g_s \ll 1$ globally
on a base $B$. In \S\ref{sec:toric} we determine the geometric
conditions for these weak coupling limits in a large ensemble of toric
bases, and find that these conditions are almost never satisfied. In
\S\ref{sec:alg} we extend this result to more general bases,
constructed via gluing local patches together, where the local patches
are crepant resolutions of orbifold singularities. In \S\ref{sec:dis}
we conclude.

\section{Non-Trivial Weak Coupling Limits}\label{sec:weakcup}
In this section we review the Sen limit and introduce a stricter limit where no $\mathcal{O}(1)$ $g_s$ regions occur. The Sen limit is the most well-known method of recovering weakly coupled type IIB from F-theory, by taking the string coupling to be perturbatively small~\cite{Sen:1996vd, Sen:1997bp}. We consider an F-theory compactification on an elliptically fibered Calabi-Yau fourfold $\pi \colon X \to B$ with section over a smooth algebraic threefold base $B$.
By a result of Nakayama~\cite{Nakayama} such an elliptic fibration
is birational to a Weierstrass model, which is described by the
equation
\begin{equation}
y^{2} = x^{3} + fx + g\, 
\end{equation}
where $f \in \Gamma(\mathcal{O}_{B}(-4K_{B}))$, $g \in \Gamma(\mathcal{O}_{B}(-6K_{B}))$ are generic sections of appropriate powers of the anticanonical bundle on $B$. The corresponding discriminant locus is given by the vanishing of
\begin{equation}
\Delta = 4f^{3} + 27g^{2}\, 
\end{equation}
which specifies the singular fibers and location of 7-branes in $B$. The complex structure modulus of the elliptic fiber corresponds to the axio-dilaton 
\begin{equation}
\tau = C_{0} + ie^{-\phi}\, 
\end{equation}which varies over the base $B$. Here the vacuum expectation value of $\phi$ determines the string coupling constant as $g_s  = e^{\phi}$. The $J$-invariant, defined in terms of the usual $j$-invariant as $J := j/1728$, takes the form
\begin{equation}
J = \frac{4f^{3}}{4f^{3} + 27g^{2}}\, 
\end{equation}
As we are interested in weak coupling we need to determine the value of the axio-dilaton from the $J$-invariant. Defining  $q = e^{2\pi i \tau}$, at large $|q|$ the $j$-invariant admits the expansion
\begin{equation}
J \simeq \frac{1}{q} + 744 + 196884 q + \dots
\end{equation}
In particular, for small $|q|$ we have that $J \sim \frac{1}{q}$ where $|q| = exp(-\frac{2\pi}{g_{s}})$, and hence studying the weak coupling limit is equivalent to studying the conditions under which $J \to \infty$. One can solve for $g_s$ for each Kodaira singular fiber by inverting the $J$-function to solve for $\tau$; see e.g.~\cite{Halverson:2016vwx}. $\tau$ is defined only up to an $SL(2,\mathbb{Z})$ action, and so $g_s$ is not uniquely defined. However, each Kodaira singular fiber has a minimal associated $g_s$ that cannot be further lowered by an $SL(2,\mathbb{Z})$ transformation. This information is presented in Table~\ref{tab:gauge}. In this sense, each 7-brane locus corresponding to a Kodaira singular fiber in Table~\ref{tab:gauge} can be thought of as a boundary condition for the axio-dilaton in the internal space, where for a fixed $SL(2,\mathbb{Z})$ frame a 7-brane along a divisor $\Sigma$ uniquely specifies the axio-dilaton along $\Sigma$. Away from these 7-brane loci the axiodilaton depends on complex structure moduli, and can be fixed by flux.

\begin{table}[t]
\begin{center}
\scalebox{.9}{\begin{tabular}{|c|c|c|c|c|c|c|c|c}
\hline
$F_i$ & $l_i$ & $m_i$ & $n_i$ & Sing. & $G_i$ & $\tau$ & $g_s$ \\  \hline
$I_0$&$\geq $ 0 & $\geq $ 0 & 0 & none & none & $\mathbb{H}$& $\geq 0$ \\
$I_n$ &0 & 0 & $n \geq 2$ & $A_{n-1}$ & $\gsu(n)$ or $\gsp(\lfloor
n/2\rfloor)$ & $i\infty$ & $0$\\
$II$ & $\geq 1$ & 1 & 2 & none & none & $e^{2\pi i/3}$ & $2/\sqrt{3}$\\
$III$ &1 & $\geq 2$ &3 & $A_1$ & $\gsu(2)$ & $i$ & $1$\\
$IV$ & $\geq 2$ & 2 & 4 & $A_2$ & $\gsu(3)$  or $\gsu(2)$ & $e^{2\pi i/3}$ & $2/\sqrt{3}$\\
$I_0^*$&
$\geq 2$ & $\geq 3$ & $6$ &$D_{4}$ & $\gso(8)$ or $\gso(7)$ or $\gg_2$& $\mathbb{H}$& $\geq 0$ \\
$I_n^*$&
2 & 3 & $n \geq 7$ & $D_{n -2}$ & $\gso(2n-4)$  or $\gso(2n -5)$ & $i\infty$ & $0$\\
$IV^*$& $\geq 3$ & 4 & 8 & $E_6$ & $\ge_6$  or $\gf_4$  & $e^{2\pi i/3}$ & $2/\sqrt{3}$\\
$III^*$&3 & $\geq 5$ & 9 & $E_7$ & $\ge_7$ & $i$ & $1$\\
$II^*$& $\geq 4$ & 5 & 10 & $E_8$ & $\ge_8$  & $e^{2\pi i/3}$ & $2/\sqrt{3}$ \\ \hline
\end{tabular}}
\caption{Kodaira fiber $F_i$, singularity, and gauge group $G_i$ on
the seven-brane at $x_i=0$ for given $l_i$, $m_i$, and $n_i$. In the second last two columns we display the minimal $g_s$ with corresponding $\tau$.}
\label{tab:gauge}
\end{center}
\end{table}

From Table~\ref{tab:gauge} we can see that the singularities that allow for weak coupling are the $I_n$ series, the $I_{n}^{*}$ series, and $I_0^*$, which realize $\gsu(n)$ or $\gsp(\lfloor n/2\rfloor)$, $\gso(2n-4)$  or $\gso(2n -5)$, and $SO(8)$ or $SO(7)$ or $G_2$ gauge groups, respectively. In fact, it is precisely the  $\gsu(n)$, $\gsp(\lfloor n/2\rfloor)$, $\gso(2n-4)$, and $SO(8)$ gauge groups that are realizable in weakly-coupled IIB orientifold theories, and so we should expect the F-theory lift to reflect this somehow.

\subsection{The Sen limit}

Sen's orientifold limit is achieved by taking the following ansatz
\begin{align}
\begin{aligned}
f &= -3h^{2} + \epsilon\eta\\
g &= -2h^{3} + \epsilon h\eta - \frac{\epsilon^{2}\chi}{12}\, 
\end{aligned}
\end{align}
The J-invariant is then given by
\begin{equation}
J=-\frac{64 \left(3 h^2-\eta  \epsilon
   \right)^3}{\epsilon ^2 \left(144 h^3 \chi -144
   \eta ^2 h^2-72 \eta  h \chi  \epsilon +3 \chi
   ^2 \epsilon ^2+64 \eta ^3 \epsilon \right)}
\end{equation}
The Sen limit is the limit in which $\epsilon$ is taken to be small, in which case the $J$-invariant at leading order in $\epsilon$ is given by
\begin{equation}
J = \frac{h^{4}}{\epsilon^{2}(\eta^{2} - h\chi)}\, 
\end{equation}
It is important to note that we wish to carefully distinguish between the Sen limit, and what we will call the \textit{strict Sen limit}, in which $\epsilon \rightarrow 0$, strictly. The latter limit takes the string coupling strictly to zero away from any 7-branes, while the former allows for small but non-zero string coupling away from 7-branes. However, even if we take $\epsilon$ infinitesimally small, but non-zero, we have $O(1)$ $g_{s}$ regions along the loci where $f=0$ or $g=0$. These sit between components of the residual $I_1$ locus, part of which collapses to an $O7$-plane in the limit $\epsilon \rightarrow 0$. Thus, in the Sen limit, we may still have strongly coupled regions along certain loci in the base. In the strict limit the resulting type IIB Calabi-Yau may be identified as a double covering of the base $B$. 

In fact, particular 7-branes known as non-Higgsable 7-branes (NH7) play a major role in obstructing weak coupling limits. NH7s arise along a divisor $D$ when $f$ and $g$ vanish to a minimal non-zero order along $D$, for all choices of complex structure moduli. This implies we can write $f = z^n \tilde{f}$ and $g = z^m \tilde{g}$, where both $n$ and $m$ are greater than zero. A key point is that $I_1$ and $A_n$ fibers cannot arise on NH7s, as the multiplicities of vanishing of $f$ and $g$ are already too high, and so for a weak coupling limit to exist we must tune to $I_0^*$ or $D_n$ on any NH7s. 

Clusters of intersecting NH7s are aptly named non-Higgsable clusters (NHCs), and they have been well-studied in the literature. Some references particularly relevant for this
  paper were cited in the introduction, but we would like to briefly mention their applications in broader contexts as well. In six dimensions, NHCs are particularly well-understood, beginning with seminal first works \cite{Morrison:1996pp,Morrison:2012np,Morrison:2012js} and
also later studies \cite{Taylor:2015ppa, Morrison:2012js,Taylor:2012dr,Morrison:2014era,Martini:2014iza,Johnson:2014xpa,Taylor:2015isa} on a variety of topics. Non-Higgsable clusters allow for gauge symmetry to arise in stabilized vacua at generic points in moduli space~\cite{Grassi:2014zxa}, i.e. without tuning to high codimension in moduli space~\cite{Braun:2014xka, Watari:2015ysa, Halverson:2016tve}, and may allow for a simple realization of the Standard Model in F-theory~\cite{Grassi:2014zxa}. 4d NHCs~\cite{Morrison:2014lca} are particularly interesting due to the fact that they can realize structure not possible in high dimensions, such as loops~\cite{Morrison:2012np}. NHCs appear in the geometry with the most number of flux vacua known~\cite{Taylor:2015xtz}, as well as in large ensembles of geometries~\cite{Halverson:2015jua}. Recently aspects of NHCs have been studied using supervised machine learning~\cite{Carifio:2017bov}.

Let us understand this from the perspective of the $J$-invariant.  The minimal non-Higgsable 7-brane corresponds to the type $II$ fiber, so suppose that we start with a divisor with a type $II$ fiber above it. Let the 7-brane locus be given by the vanishing of a local coordinate $z$. Near $z = 0$ we can Taylor expand $f$ and $g$ as
\begin{align}
& f = f_0 + f_1 z + f_2 z^2 + f_3 z^3 + \dots\,  \nonumber \\
& g = g_0 + g_1 z + g_2 z^2 + g_3 z^3 + \dots\, 
\end{align}
For a type $II$ fiber we have $f_0 = g_0 = 0$, and so we can write $f = z \tilde{f}$ and $g = z \tilde{g}$. The $J$-invariant thus takes the form, 
\begin{equation}
J = \frac{4z\tilde{f}^3}{4z\tilde{f}^3 + 27 \tilde{g}^2}\, 
\end{equation}
This has strong coupling points along $z = 0$. In order to get rid of the strong coupling point at $z = 0$ one can try to tune to remove the factors of $z$. The only way to do so is to take $g = z^2 \tilde{g}$, which gives a type $III$ fiber. However, this leaves us with
\begin{equation}
J = \frac{4\tilde{f}^3}{4\tilde{f}^3 + 27 z\tilde{g}^2}\, 
\end{equation}
so we are again left with strong coupling along $z = 0$. One can continue on in the same fashion, until we hit the $f = z^2 \tilde{f}$, $g = z^3 \tilde{g}$, which is the case of $I_0^*$. Here the $J$-invariant reads
\begin{equation}
J = \frac{4\tilde{f}^3}{4\tilde{f}^3 + 27 \tilde{g}^2}\, 
\end{equation}
which does not necessarily have strong coupling along $z= 0$. However, if we start with a type $IV^*$ fiber it is easy to see in a similar manner that one cannot tune $f$ and $g$ to eliminate all strong coupling points on 7-branes simultaneously. Therefore we confirm that the singular fibers with possible Sen limits, where there are no regions of $\mathcal{O}(1)\, g_s$ along 7-branes that carry a gauge group, are $I_n$, $I_n^{*}$, and $I_0^*$.

One feature that distinguishes the Sen limit from the strict Sen limit is the allowed gauge groups, as a monodromy action on the fibers can reduce the rank of the gauge group. In the strict Sen limit the gauge group $G_2$ is not present, as the monodromy ramification locus necessary to reduce $SO(8)$ to $G_2$ disappears as $\epsilon \rightarrow 0$~\cite{Blumenhagen:2009up}. However, for finite but small $\epsilon$ this locus can still be present, and $G_2$ can still be realized.

If one insists on the stronger condition of a global weak-coupling limit, without any regions of $\mathcal{O}(1)$ $g_s$, the 7-brane charge needs to be cancelled everywhere locally, which restricts us to a fiber of type $I_0^{*}$. In analogy to the Sen limit we now define the \textit{global weak coupling limit}.

\subsection{The Global Weak Coupling Limit}

We say that a smooth algebraic threefold base $B$ admits a global weak coupling limit (GWCL), if the corresponding Weierstrass model $\pi \colon X \to B$ can be tuned so that  $g_{s} \ll 1$ globally on the base $B$. As shown in e.g~\cite{Halverson:2016vwx}, a necessary condition to achieve weak coupling everywhere in $B$ is to have an $I_{0}^{*}$ fiber on all non-Higgsable seven branes. Indeed, in the most generic such case, $f$ and $g$ take the following form
\begin{equation}
f = Fm^{2}  \qquad g = Gm^{3}\, 
\end{equation}
where $m \in \Gamma(\mathcal{O_{B}}(-2K_{B}))$. Here $\,F,G \in \Gamma(\mathcal{O_{B}})$ can be thought of as complex constants and we have non-Higgsable 7-branes along the vanishing loci of $m$. Moreover, $m$ should be reduced, possibly with normal crossing singularities. Given such a parametrization, the J-invariant thus takes the form
\begin{equation}
J = \frac{4F^{3}}{4F^{3} + 27G^{2}}\,
\end{equation}
which is indeed constant everywhere on the base. Thus, in order to achieve weak coupling, we must take the expression $4F^{3} + 27G^{2}$ sufficiently small.

It suffices to parametrize $F$ and $G$ appropriately to take $J$ to be arbitrarily large. Let us take the following ansatz for the constants $F$ and $G$
\begin{align}
\begin{aligned}
F &\coloneqq -3\,  \\
G &\coloneqq -2 + \tilde{\epsilon}\, 
\end{aligned}
\end{align}
where $\tilde{\epsilon}$ is an arbitrary parameter. At small $\tilde{\epsilon}$ we therefore have
\begin{equation}
J = \frac{1}{\tilde{\epsilon}}\, 
\end{equation}
We thus obtain the weak coupling limit by taking the parameter $\tilde{\epsilon} \to 0$, in other words, taking $G \to -2$.

We may thus generalize the Sen parametrization by taking the following ansatz
\begin{align}
f &= -3h^{2} + \epsilon\eta\,  \\
g &= (-2 + \tilde{\epsilon})\,h^{3} + \epsilon h\eta - \frac{\epsilon^{2}\chi}{12}\, 
\end{align}
where we simply take the transformation $-2 \mapsto -2 + \tilde{\epsilon}$ in the original Sen parametrization. By taking the limit $\tilde{\epsilon} \to 0$, we recover Sen's original expressions, while successively taking the limits $\epsilon \to 0$ and then $\tilde{\epsilon} \to 0$ recovers our original ansatz and allows us to achieve weak coupling everywhere in the base. 

For a GWCL, any seven-brane must have an associated singular fiber that is $I_0^*$, but in general this does not uniquely determine the gauge group, due to monodromy effects. In the case of an $I_0^*$ fiber along a locus $\Sigma$ the gauge group is determined from a monodromy-cover $\tilde{\Sigma}$ of $\Sigma$, which takes the form $\psi^3 + (f/z^2)|_{z = 0} \psi + (g/z^3)|_{z = 0}$, where $\psi$ is a generic section of a particular line bundle on $X$~\cite{Grassi:2011hq}. If $F,G\in \Gamma(\mathcal{O_{B}})$ then the monodromy-cover has no monodromy, and the gauge group is necessarily $SO(8)$, as opposed to $SO(7)$ or $G_2$, which are also possible with an $I_0^*$ fiber. This is true even in the presence of multiple $I_0^*$ loci, as one can explicitly check that each Dynkin node is invariant under any possible monodromy action.

We now turn to studying the geometric requirements for both of these weak coupling limits. We first examine this in the toric setting, from which most examples so far have been drawn.

\section{ Weak Coupling Limits on Toric Bases}\label{sec:toric}
In this section we derive rather constraining requirements for a toric base to admit either of the two weak coupling limits. In~\cite{Halverson:2017ffz} an ensemble of $\frac43 \times 2.96\times 10^{755}$ toric bases for F-theory was generated by crepant fourfold transitions induced by base changes from certain minimal geometries that we will discuss. We will discuss both general toric bases and those that are part of this ensemble. We will begin by reviewing the construction of the ensemble and then proceed by analyzing the GWCL first as it is more restrictive and therefore simpler than the Sen's limit.

We first review some basic facts about toric varieties. Let $B$ be a smooth toric variety, corresponding to a fan $\Sigma$, and denote the rays of $\Sigma$ as $u_i \in \Sigma(1)$. Writing $D = \sum_i a_i D_i$, where the $D_i$ are the toric divisors corresponding to the rays $u_i$, global sections $\Gamma( \mathcal{O}(D))$ are given by points $m$ in a lattice denoted $M$ such that $ m \cdot u_i  \geq -a_i$ for all $i$. In particular, global sections of $\mathcal{O}(-6 K_B)$ are given by $m$ such that $ m \cdot u_i  \geq -6$ for all $i$, and global sections of $\mathcal{O}(-4 K_B)$ are given by $m$ such that $ m \cdot u_i  \geq -4$. The $m$'s that satisfy these conditions form lattice polytopes, which we denote $\Delta_g$ and $\Delta_f$:
\begin{equation}
\Delta_f = \{m\in \bZ^3 \, | \, m\cdot v_i + 4 \geq 0 \,\, \forall\, i \} \, \qquad \quad
\Delta_g = \{m\in \bZ^3 \, | \, m\cdot v_i + 6 \geq 0 \,\, \forall \, i \}\, 
\label{eqn:dfdg}
\end{equation}
which correspond to monomials via
\begin{equation}\label{eqn:mons}
m_{f} \in \Delta_{f} \mapsto \prod_{i}x_{i}^{m_{f}\cdot v_{i} + 4}\,  \qquad \quad
m_{g} \in \Delta_{g} \mapsto \prod_{i}x_{i}^{m_{g}\cdot v_{i} + 6}\, 
\end{equation}
where each $x_{i}$ is a homogenous coordinate on $B$ corresponding to each $v_{i}$. The most generic forms for the sections $f$ and $g$ thus take the following form
\begin{equation}
f = \sum_{m_{f} \in \Delta_{f}}a_{f} \prod_{i}x_{i}^{m_{f}\cdot v_{i} + 4}\,  \qquad \quad
g = \sum_{m_{g} \in \Delta_{g}}a_{g} \prod_{i} x_{i}^{m_{g} \cdot v_{i} + 6}\, 
\end{equation}
with the $a_f$ and $a_g$ generic complex coefficients. 
In the following we will be interested in generating a large ensemble of geometries from crepant base-changing resolutions of some minimal geometries. The minimal geometries we consider are smooth weak Fano toric varieties associated with a fine regular star triangulation (FRST) of a 3d reflexive polytope $\Delta^\circ$, with corresponding face fan $\Sigma$. Of key interest in this work will be the Cox ring, or the homogenous coordinate ring, of the toric variety; in particular, the behavior of sections of line bundles under a crepant resolution. Each of these toric varieties can be viewed as a crepant resolution of a singular toric variety $\tilde{B}$, whose coordinate ring is generated by the corners (vertices) of $\Delta^\circ$. This toric variety may not even be simplicial, though one can make it so by appropriately subdividing the associated fan. Resolving $\tilde{B}$ by triangulating $\Delta^\circ$ introduces exceptional toric divisors, that correspond to points interior to edges and facets of $\Delta^\circ$. These divisors are rigid, in the sense that they have no normal bundle deformations. This structure will be important both in determining if there are non-Higgsable clusters, and the presence of either weak coupling limit.

On a toric base it will often be the case that a single base-changing resolution introduces at least a type $II$ fiber on one of these crepant exceptional divisors $D$. Therefore, to realize either of the weak coupling limits we must tune to an $I_n^*$ fiber for $n \geq 0$, as the type $II$ non-Higgsable 7-branes eliminate the possibility of an $I_n$ fiber\footnote{$A_1$ and $A_2$ gauge
algebras may arise from type $III$ and $IV$ Kodaira fibers, but these
do not exist at weak coupling. See \cite{Grassi:2014zxa,Halverson:2016vwx} for related discussion.}. This requires $f$ and $g$ to factorize as
\begin{equation}
f = c_f F \prod_i s_i^2\, \quad
g = c_g G \prod_i s_i^3\,
\end{equation}
where the $s_i$ are irreducible polynomials that satisfy $\sum_i 2 [s_i] - F = -4 K_B$,\, $\sum_i 3 [s_i] - G = -6 K_B $. $F$ and $G$ are non-trivial polynomials that parametrize a possible $I_1$ locus, and $c_f$ and $c_g$ are constants. However, as a single blowup generically forces type $II$ on many toric divisors, we will be mainly interested in the case that the $s_i$ correspond to toric coordinates $x_i$. 

In the following we will find that the existence of a generic toric blowup on a toric base $B$ naturally induces type II fibers on nearby crepant exceptional divisors. As singular fibers can only be enhanced, we must engineer transitions to tune to singular fibers supporting a GWCL or Sen limit along the starting type II fibers. It then follows, via simple combinatorics and using necessary conditions on orders of vanishing for transitions in the Weierstrass model, that such bases supporting either limit are scarce, and always admit multiple transitions to bases prohibiting either limit. On the level of the geometry, as we blowup and move away from the Fano range and simultaneously increase the Kodaira dimension, i.e. as the number of sections of the canonical bundle grow, we can image the space of sections of the anticanonical bundle gradually shrinking, and its base locus, i.e. the common locus of all the sections, enlarging, while the multiplicity of the sections increases simultaneously. Such a base locus will typically correspond to rigid crepant exceptional divisors, and therefore the existence of rigid crepant exceptional divisors is crucial to an obstruction of either of the weak coupling limits. Such exceptional divisors are the easiest to move within the base locus and to augment its multiplicity past the orders required for weak coupling limits.

\subsection{Trees of F-theory Geometries}
In this section we give a brief overview of the ensemble of F-theory geometries constructed in~\cite{Halverson:2017ffz}. This is the largest ensemble of Calabi-Yau fourfolds known, and the scarcity of weak coupling limits in this ensemble provides strong evidence against the existence of weak coupling limits for a generic Calabi-Yau fourfold. We will henceforth refer to the corresponding bases of the elliptic fibration as toric tree bases.

\begin{figure}
\begin{center} 
\begin{tikzpicture}[scale=3]
 \draw[thick,color=Black] (.7101,.5) --
(-.71,.5);\draw[thick,dash pattern={on 1pt off 1pt},color=ForestGreen] (-.71,.5)--(-.51,1.5)--(0,1)--(.5101,1.5)--(.7101,.5); \fill (0,0) circle (.5mm); \fill (.7101,.5) circle (.5mm); \fill
(-.71,.5) circle (.5mm); \fill (0,1) circle (.5mm); \fill (-.51,1.5) circle
(.5mm); \fill (.5101,1.5) circle (.5mm); \draw (0,0) -- (.7101,.5); \draw (0,0) --
(-.71,.5); \draw (0,0) -- (0,1); \draw (0,0) -- (-.51,1.5); \draw (0,0) --
(.5101,1.5); \node at (-.89,.5) {$v_1$}; \node at (.89,.5) {$v_2$}; \node at
(0,-.21) {$0$}; \node at (0,1.151) {$2$}; \node at (-.62,1.6) {$3$}; \node at
(.62,1.6) {$3$}; 
\end{tikzpicture}
  \caption{\label{fig:smalltree} A height-2 and two height-3 blowups of an edge generated by the vertices $\{v_1, v_2\}$.}
\end{center}
\end{figure}
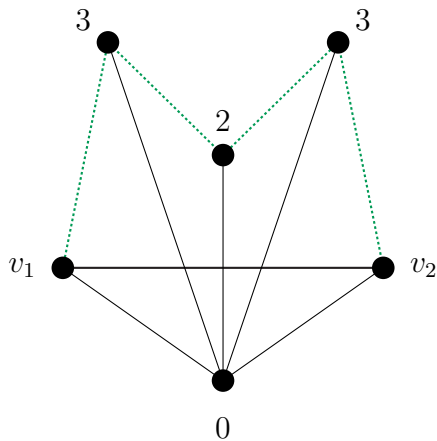

Let $B$ be a smooth weak Fano toric threefold induced by a crepant resolution of a Fano toric threefold corresponding to a 3d reflexive polytope $\Delta^{\circ}$ via an FRST. Associated to such a $B$ is a fan consisting of 2d and 3d cones corresponding to edges and faces on facets of $\Delta^{\circ}$. For a fixed facet, the number of such edges and faces is completely determined by the number of vertices and boundary lattice points interior to each edge which is independent of the choice of triangulation. 
Beginning from a fixed weak Fano toric threefold, transitions to topologically distinct threefolds will be obtained via smooth toric blowups of curves or points. These birational morphisms can be interpreted purely combinatorially at the level of polytopes. For instance, to blowup a curve, given 2 cones with generators $(v_{0}, v_{1}, v_{2})$ and $(v_{1}, v_{2}, v_{3})$ we add an additional ray $v_{e} \coloneqq v_{1} + v_{2}$ which replaces the original 2 cones with 4 cones given by the generators $(v_{0}, v_{1}, v_{e}), (v_{0}, v_{2}, v_{e}), (v_{1}, v_{e}, v_{3}), (v_{2}, v_{e}, v_{3})$. We can repeat this procedure by recursively subdividing, which yields the diagram shown in Fig.~\ref{fig:smalltree} in two dimensions,
where the edge between $v_{1}$ and $v_2$ corresponds to an edge contained in a facet, and we subdivide this edge by adding additional rays. In particular, we will refer to the cones corresponding to a sequence of blowups as illustrated above colloquially as ``trees'' which is further emphasized from the green dashed lines corresponding to new edges resulting from these subdivisions. From such a procedure, it is clear that any additional ray $v_{e}$ must take the form $v_{e} = \sum_{i}a_{i}v_{i}$ which is a linear combination of two vertices if it lies over an edge, or three vertices if it lies over a face, where each $v_{i}$ is a lattice point of $B_i$. Such additional rays will also be informally called ``leaves'' with ``height'' $h$ defined by $h = \sum_{i}a_{i}$ where the heights of the leaves are labeled in the above diagram. The leaves with $h = 1$ correspond to lattice points on the facet which we will refer to as ``roots''. Given a tree obtained from a sequence of subdivisions, we will refer to the height of the tree as the height of a highest leaf. 

Instead of considering all possible trees above a fixed facet, which is computationally infeasible, we will consider trees built above individual simplices on the facet. Thus, we will refer to trees above a fixed face or edge as face trees or edge trees, respectively. Given a fixed height, all trees with height at most $h$ can be classified purely via combinatorial techniques. In particular, to visualize such a combinatorial procedure, it is fruitful to picture each tree with its leaves projected to the base edge or face. To illustrate this, given a ground edge
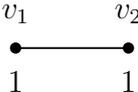
\begin{center}
\begin{tikzpicture}[scale=1.5]
\draw[thick,color=Black] (0,0) -- (1,0);
\fill (0,0) circle (.5mm);
\fill (1,0) circle (.5mm);
\node at (0,.3) {$v_1$};
\node at (1,.3) {$v_2$};
\node at (0,-.3) {$1$};
\node at (1,-.3) {$1$};
\end{tikzpicture}
\end{center}
we may subdivide by adding a point on the edge and its corresponding height as pictured
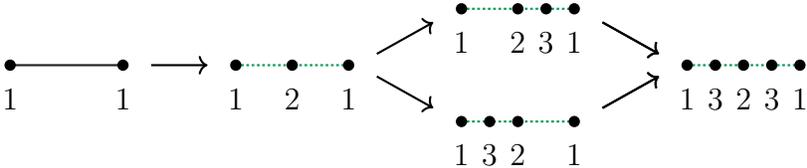
\begin{center}
\begin{tikzpicture}[scale=1.5]
\draw[thick,color=Black] (0,0) -- (1,0);
\fill (0,0) circle (.5mm);
\fill (1,0) circle (.5mm);
\node at (0,-.3) {$1$};
\node at (1,-.3) {$1$};
\draw[thick,->] (1.25,0) -- (1.75,0);
\draw[thick,dash pattern={on 1pt off 1pt},color=ForestGreen] (2,0) -- (3,0);
\fill (2,0) circle (.5mm);
\node at (2,-.3) {$1$};
\fill (2.5,0) circle (.5mm);
\node at (2.5,-.3) {$2$};
\fill (3,0) circle (.5mm);
\node at (3,-.3) {$1$};
\draw[thick,->] (3.25,.1) -- (3.75,.38);
\draw[thick,->] (3.25,-.1) -- (3.75,-.38);
\draw[thick,->] (5.25,.38) -- (5.75,.1);
\draw[thick,->] (5.25,-.38) -- (5.75,-.1);
\draw[thick,dash pattern={on 1pt off 1pt},color=ForestGreen] (4,.5) -- (5,.5);
\fill (4,.0+.5) circle (.5mm);
\node at (4,-.3+.5) {$1$};
\fill (4.5,0+.5) circle (.5mm);
\node at (4.5,-.3+.5) {$2$};
\fill (4.75,0+.5) circle (.5mm);
\node at (4.75,-.3+.5) {$3$};
\fill (5,0+.5) circle (.5mm);
\node at (5,-.3+.5) {$1$};
\draw[thick,dash pattern={on 1pt off 1pt},color=ForestGreen] (4,-.5) -- (5,-.5);
\fill (4,.0-.5) circle (.5mm);
\node at (4,-.3-.5) {$1$};
\fill (4.5,0-.5) circle (.5mm);
\node at (4.5,-.3-.5) {$2$};
\fill (4.25,0-.5) circle (.5mm);
\node at (4.25,-.3-.5) {$3$};
\fill (5,0-.5) circle (.5mm);
\node at (5,-.3-.5) {$1$};
\draw[thick,->] (5.25,.38) -- (5.75,.1);
\draw[thick,->] (5.25,-.38) -- (5.75,-.1);
\draw[thick,dash pattern={on 1pt off 1pt},color=ForestGreen] (6,0) -- (7,0);
\fill (6,.0) circle (.5mm);
\node at (6,-.3) {$1$};
\fill (6.25,0) circle (.5mm);
\node at (6.25,-.3) {$3$};
\fill (6.5,0) circle (.5mm);
\node at (6.5,-.3) {$2$};
\fill (6.75,0) circle (.5mm);
\node at (6.75,-.3) {$3$};
\fill (7,0) circle (.5mm);
\node at (7,-.3) {$1$};
\end{tikzpicture}
\end{center}
where the last subdivided edge is the above tree with all its leaves projected down. In a completely analogous fashion, a subdivision of a face may be pictured as follows
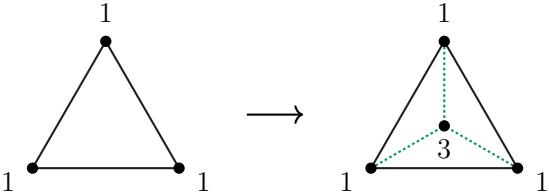
\begin{center}
\begin{tikzpicture}[scale=1.5, every node/.style={scale=0.9}]
\draw[thick,color=Black] (90:.75) -- (90+120:.75) -- (90+120+120:.75) -- cycle;
\fill (90:.75) circle (.5mm);
\fill (90+120:.75) circle (.5mm);
\fill (90+240:.75) circle (.5mm);
\node at (90:1) {$1$}; \node at (90+120:1) {$1$}; \node at (90+240:1) {$1$};
\draw[thick,->] (1.25,.1) -- (1.75,.1);
\begin{scope}[xshift=3cm]
\draw[thick,dash pattern={on 1pt off 1pt},color=ForestGreen] (90:.75) -- (0,0);
\draw[thick,dash pattern={on 1pt off 1pt},color=ForestGreen] (90+120:.75) -- (0,0);
\draw[thick,dash pattern={on 1pt off 1pt},color=ForestGreen] (90+240:.75) -- (0,0);
\fill (0,0) circle (.5mm);
\draw[thick,color=Black] (90:.75) -- (90+120:.75) -- (90+120+120:.75) -- cycle;
\fill (90:.75) circle (.5mm);
\fill (90+120:.75) circle (.5mm);
\fill (90+240:.75) circle (.5mm);
\node at (90:1) {$1$}; \node at (90+120:1) {$1$}; \node at (90+240:1) {$1$};
\node at (0,-.2) {$3$};
\end{scope}
\end{tikzpicture}
\end{center}
where the additional point with height $3$ corresponds to adding a ray by summing all three of the roots. 

There is a strong constraint on the full moduli space of elliptically fibered Calabi-Yau fourfolds which give consistent F-theory compactifications, namely the exclusion of fourfolds containing $(4,6)$ divisors. It is easy to see, via a completely combinatorial argument~\cite{Halverson:2017ffz}, that this $(4,6)$ condition is satisfied by imposing an upper bound $h \leq 6$ on the height of all trees. Thus, for all heights $3 \leq h \leq 6$, we may enumerate all possible trees which gives the numbers in Table~\ref{tab:heights}.
\begin{table}
\begin{center}
\begin{tabular}{|c|c|c|}
\hline
$N$ & \# Edge Trees & \# Face Trees \\ \hline
$3$ & $5$ & $2$\\
$4$ & $10$ & $17$\\
$5$ & $50$ & $4231$ \\
$6$ & $82$ & $41,873,645$\\ \hline
\end{tabular}
\caption{\label{tab:heights} The number of possible edge trees and face trees as a function of the maximal height $h$.}
\end{center}
\end{table}

Given a 3d reflexive polytope, we may define a corresponding ensemble $S_{\Delta^{\circ}}$. Let $\mathcal{T}(\Delta^{\circ})$ be an FRST of $\Delta^{\circ}$. For each face or edge, we add a face or edge tree with maximal height $6$. Then, from the above table, one finds that the number of possibilities in the ensemble $S_{\Delta^{\circ}}$ is 
\begin{equation}
|S_{\Delta^{\circ}}| = 82^{\#\tilde{E}\text{ on }\mathcal{T}(\Delta^{\circ})}\times (41,873,645)^{\#\tilde{F}\text{ on }\mathcal{T}(\Delta^{\circ})}\, ,
\end{equation}
where we define $\#\tilde{E}$ and $\#\tilde{F}$ as the number of total edges and faces on $\mathcal{T}(\Delta^{\circ})$.

\begin{figure}
\begin{center}
\begin{tikzpicture}[scale=1.7]
\draw[thick,color=Black] (0,0) -- (3,0) -- (0,3) -- cycle;
\draw[thick,color=Black] (0,.5) -- (2.5,.5);
\draw[thick,color=Black] (0,1) -- (2,1);
\draw[thick,color=Black] (0,1.5) -- (1.5,1.5);
\draw[thick,color=Black] (0,2) -- (1,2);
\draw[thick,color=Black] (0,2.5) -- (.5,2.5);
\draw[thick,color=Black] (.5,0) -- (.5,2.5);
\draw[thick,color=Black] (1,0) -- (1,2);
\draw[thick,color=Black] (1.5,0) -- (1.5,1.5);
\draw[thick,color=Black] (2,0) -- (2,1);
\draw[thick,color=Black] (2.5,0) -- (2.5,.5);
\draw[thick,color=Black] (0,2) -- (.5,2.5);
\draw[thick,color=Black] (0,1) -- (1,2);
\draw[thick,color=Black] (0,0) -- (1.5,1.5);
\draw[thick,color=Black] (1,0) -- (2,1);
\draw[thick,color=Black] (2,0) -- (2.5,.5);
\draw[thick,color=Black] (0,1.5) -- (.5,2);
\draw[thick,color=Black] (0,.5) -- (1,1.5);
\draw[thick,color=Black] (.5,0) -- (1.5,1);
\draw[thick,color=Black] (1.5,0) -- (2,.5);
\fill (0,0) circle (.5mm); \fill (0,.5) circle (.5mm); \fill (0,1) circle (.5mm);
\fill (0,1.5) circle (.5mm); \fill (0,2) circle (.5mm); \fill (0,2.5) circle (.5mm);
\fill (0,3) circle (.5mm);
\fill (.5,0) circle (.5mm); \fill (.5,.5) circle (.5mm); \fill (.5,1) circle (.5mm);
\fill (.5,1.5) circle (.5mm); \fill (.5,2) circle (.5mm); \fill (.5,2.5) circle (.5mm);
\fill (1,0) circle (.5mm); \fill (1,.5) circle (.5mm); \fill (1,1) circle (.5mm);
\fill (1,1.5) circle (.5mm); \fill (1,2) circle (.5mm); 
\fill (1.5,0) circle (.5mm); \fill (1.5,.5) circle (.5mm); \fill (1.5,1) circle (.5mm);
\fill (1.5,1.5) circle (.5mm); 
\fill (2,0) circle (.5mm); \fill (2,.5) circle (.5mm); \fill (2,1) circle (.5mm);
\fill (2.5,0) circle (.5mm); \fill (2.5,.5) circle (.5mm);
\fill (3,0) circle (.5mm);
\draw[thick,color=Black] (1,3) -- (7,3) -- (7,0) -- cycle;
\draw[thick,color=Black] (3,2) -- (7,2);
\draw[thick,color=Black] (5,1) -- (7,1);
\draw[thick,color=Black] (6.5,1) -- (6.5,3);
\draw[thick,color=Black] (6,1) -- (6,3);
\draw[thick,color=Black] (5.5,1) -- (5.5,3);
\draw[thick,color=Black] (5,1) -- (5,3);
\draw[thick,color=Black] (4.5,2) -- (4.5,3);
\draw[thick,color=Black] (4,2) -- (4,3);
\draw[thick,color=Black] (3.5,2) -- (3.5,3);
\draw[thick,color=Black] (3,2) -- (3,3);
\draw[thick,color=Black] (7,1) -- (6,3);
\draw[thick,color=Black] (6.5,1) -- (5.5,3);
\draw[thick,color=Black] (6,1) -- (5,3);
\draw[thick,color=Black] (5.5,1) -- (4.5,3);
\draw[thick,color=Black] (5,1) -- (4,3);
\draw[thick,color=Black] (7,2) -- (6.5,3);
\draw[thick,color=Black] (4,2) -- (3.5,3);
\draw[thick,color=Black] (3.5,2) -- (3,3);
\draw[thick,color=Black] (7,0) -- (6.5,1);
\draw[thick,color=Black] (7,0) -- (6,1);
\draw[thick,color=Black] (7,0) -- (5.5,1);
\draw[thick,color=Black] (5,1) -- (4.5,2);
\draw[thick,color=Black] (5,1) -- (4,2);
\draw[thick,color=Black] (5,1) -- (3.5,2);
\draw[thick,color=Black] (3,2) -- (2.5,3);
\draw[thick,color=Black] (3,2) -- (2,3);
\draw[thick,color=Black] (3,2) -- (1.5,3);
\fill (1,3) circle (.5mm); \fill (1.5,3) circle (.5mm); \fill (2,3) circle (.5mm); \fill (2.5,3) circle (.5mm);
\fill (3,3) circle (.5mm); \fill (3.5,3) circle (.5mm); \fill (4,3) circle (.5mm); \fill (4.5,3) circle (.5mm);
\fill (5,3) circle (.5mm); \fill (5.5,3) circle (.5mm); \fill (6,3) circle (.5mm); \fill (6.5,3) circle (.5mm);
\fill (7,3) circle (.5mm);
\fill (7,2) circle (.5mm);
\fill (7,1) circle (.5mm);
\fill (7,0) circle (.5mm);
\fill (6.5,1) circle (.5mm);
\fill (6,1) circle (.5mm);
\fill (5.5,1) circle (.5mm);
\fill (5,1) circle (.5mm);
\fill (6.5,2) circle (.5mm);
\fill (6,2) circle (.5mm);
\fill (5.5,2) circle (.5mm);
\fill (5,2) circle (.5mm);
\fill (4.5,2) circle (.5mm);
\fill (4,2) circle (.5mm);
\fill (3.5,2) circle (.5mm);
\fill (3,2) circle (.5mm);

\end{tikzpicture}
\caption{\label{fig:facets} The largest facets for $\Delta^\circ_1$ and $\Delta^\circ_2$, respectively, each with an arbitrary triangulation.}
\end{center}
\end{figure}
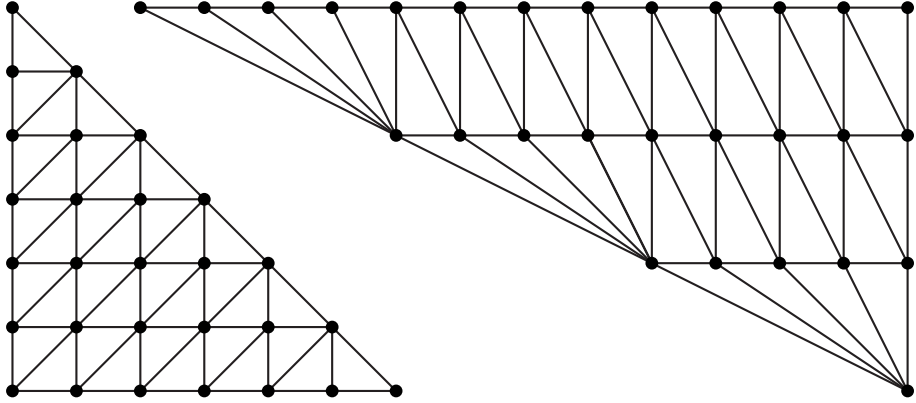

Thus, for each $\Delta^{\circ}$ of the $4,319$ reflexive polytopes, we have a corresponding ensemble $S_{\Delta^{\circ}}$. By comparing the cardinality of all $4,319$ ensembles, there are two distinct polytopes $\Delta_{1}^\circ$ and $\Delta_{2}^\circ$ which give a dominating contribution to the total number of trees. The vertex sets of these polytopes are given by
\begin{align*}
S_{1} &= \{ (-1,-1,-1),(-1,-1,5),(-1,5,-1),(1,-1,-1)\}\\
S_{2} &= \{ (-1,-1,-1),(-1,-1,11),(-1,2,-1),(1,-1,-1)\}
\end{align*}
By computing the total number of edges and faces of any triangulation for both $\Delta_{1}^\circ$ and $\Delta_{2}^\circ$, we find that they have the same number of total edges and faces given by $\# \tilde{E} = 108$ and $\# \tilde{F} = 72$. We find that the cardinality of each ensemble is given by
\begin{equation}
|S_{\Delta_{1}^{\circ}}| = \frac{2.96}{3} \times 10^{755} \qquad \qquad |S_{\Delta_{2}^{\circ}}| = 2.96 \times 10^{755}\, ,
\end{equation}
where the additional factor of $\frac{1}{3}$ in $|S_{\Delta_{1}^{\circ}}|$ corresponds to an order $3$ symmetry obtained by rotating one of the facets of $\Delta_{1}^{\circ}$. Adding these numbers gives a lower bound of $\frac{4}{3} \times 2.96 \times 10^{755}$ threefold bases supporting topologically distinct F-theory geometries. The ensemble generated by building trees over $\Delta^\circ_1$ and $\Delta^\circ_2$ completely dominate the ensemble of tree geometries. These geometries will therefore serve as the ``ground'' on which we will build our trees, yielding a vast number of topologically distinct bases for elliptically fibered Calabi-Yau fourfolds. In particular, $\Delta^\circ_1$ and $\Delta^\circ_2$ each have a facet with $63$ edges and $36$ faces, which determine much of the structure of the ensemble, including the scarcity of weak coupling limits, as we will find. The corresponding facets are shown in Fig.~\ref{fig:facets}.

\subsection{The global weak coupling limit}

As discussed above we will concentrate on the case where our $I_0^*$ loci are located on the vanishing of single toric coordinates, as opposed to the vanishing of more general polynomials thereof. This will actually provide rather strict requirements on toric bases that admit GWCLs, as blowups generically results in non-Higgsable type $II$ fibers over particular toric divisors, as we will discuss.

Since the existence of GWCL requires that the only singular fibers
are $I_0^*$ fibers, it is natural to group the toric divisors into two sets: those that support an $I_0^*$ singularity, and those that do not. Then $f$ and $g$ factorize as
\begin{align}\label{eqn:req}
f= x_1^2 \dots x_p^2 F\,  \nonumber \\
g= x_1^3 \dots x_p^3 G\, 
\end{align}
Here $F, G \in \Gamma(\mathcal{O}_B)$, and we have $I_0^*$ loci on the toric coordinates $x_1, \dots, x_p$, and smooth elliptic fibers over $x_{p+1}, \dots, x_n$. Denote the former coordinates $x_a$, and the latter $x_i$. It is simple to see that, for all $m$ corresponding to monomials in $g$, 
we have
\begin{align}
&\langle m, u_a \rangle = -3\,  \nonumber \\
 &\langle m, u_i \rangle = -6\, 
  \label{eqn:gwclcond}
\end{align}
The $-6$ condition is necessary for a smooth fiber over $D_i$ and no $I_1$
fiber anywhere, where the latter is necessary for a GWCL.

\newpage
We now consider which toric bases $B$ allow for a GWCL. We will see that under
  simple geometric conditions on $B$ the GWCL condition \ref{eqn:gwclcond} is
violated and F-theory on $B$ does not admit a GWCL. Let us take our base $B$ to be obtained by a sequence of toric blowups from one of the smooth weak Fano toric threefolds given by an FRST of a reflexive 3d polytope $\Delta^\circ$. 

First consider a point $p$ interior to a facet $F$, corresponding to a toric coordinate $x_p$ with divisor $D_p$. Assume that we have a type $II$ fiber on $D_p$, and so to realize a GWCL we need to tune an $I_0^*$ on $D_p$, then we need to include only monomials in $g$ whose corresponding $m$ (see Eq.~\ref{eqn:mons}) satisfy $ m \cdot p_f  = -3$. Suppose that $B$ indeed admits a GWCL. Then by linearity across $F$, these monomials must also satisfy $ m \cdot v  = -3$, for all $v \in F$. To see this, express $p = a_1 v_1 + a_2 v_2 + (1 -a_1 - a_2) v_3$, where $v_1, v_2, v_3$ are vertices of $F$, and $a_1, a_2 > 0$. By assumption of the existence of the GWCL we have $v_i \cdot m \in \{-3, -6\}$ for all $m \in \Delta_g $, $i = 1,2,3$, One then sees that the only solution consistent with the GWCL is that $v_i \cdot m = -3$ for all $m$, $i = 1,2,3$.

This immediately shows that if we start with a type $II$ fiber on $D_p$, we must tune an $I_0^*$ on every point on $F$ to realize a GWCL.  A similar result holds for any point $p_e$ interior to edges: if we tune an $I_0^*$ on $p_e$, then we must have $I_0^*$ on the entire edge.

Let us now start with a toric base geometry corresponding to an FRST of a 3d reflexive polytope, and consider the effect of toric blowups. Here we are simply blowing up in a coordinate patch, so that the blowup divisors are generated by two or three rays in $\Sigma$ that belong to a common cone. Each blowup ray can then be written as $v_e = \sum_i a_i u_i$.  In general, the blowing up of $B$ will reduce the number of sections of $\mathcal{O}(-4 K_B)$ and $\mathcal{O}(-6 K_B)$, since adding rays to the fan induces more hyperplane constraints on $\Delta_f$ and $\Delta_g$. Without loss of generality we will concentrate on $\Delta_g$, since the result for $\Delta_f$ is nearly identical. Assume that we have an $I_0^*$ on a point interior to a facet $F$, which implies $I_0^*$ fibers on all toric divisors corresponding to points on $F$. Now consider two intersecting toric divisors $D_1$ and $D_2$, with corresponding rays $v_1$ and $v_2$ on $F$. Let us build a tree above the edge defined by $\{ v_1, v_2\}$. There will then be corresponding divisors of the form $D = a_1 D_1 + a_2 D_2$, and we have
\begin{equation}
  mult_D(g) = \langle m, a_{1}v_{1} + a_{2}v_{2} \rangle + 6 = -3(a_1+a_2)+6= -3 h + 6 
\end{equation}
for all $m\in \Delta_g$.
Thus, for any $h \geq 3$ blowup, it must be the case that $m\notin \Delta_g$ and therefore that the monomials supporting
$I_0^*$ fibers above all points in $F$ are absent. That is, the
condition \eqref{eqn:gwclcond} is violated and the GWCL is spoiled, as the multiplicity of vanishing in $f$ and $g$ is greater than that in $I_0^{*}$; i.e. a $IV^*$, $III^*$, or $II^*$ fiber is obtained, all of which correspond to seven-branes with exceptional geometric gauge group, none of which admit a weak coupling limit. An even stricter result holds for face blowups: any face blowup necessarily has $h \geq 3$, and so any face blowup necessarily eliminates the monomials allowing for $I_0^*$ fibers, and therefore also spoils the GWCL.

 A similar result hold for a point interior to a polytope edge $e$. Let us assume we have at least a type $II$ fiber on the divisor corresponding to such a point, which implies that there are $I_0^*$ fibers on all toric divisors corresponding to points on $e$. Now consider two of these toric divisors $D_1$ and $D_2$, with corresponding rays $v_1$ and $v_2$. If we perform a blowup using these rays, such that $D = a_1 D_1 + a_2 D_2$, it is simple to see that $mult_D(g) = -3 h+6$, and therefore any $h \geq 3$ blowup will eliminate the monomials that give the correct multiplicity of vanishing for the points interior for $e$ to support $I_0^*$ fibers. Then the GWCL is spoiled.

Summarizing, we have a strong constraint on the toric tree bases that admit a GWCL: a GWCL can only exist if the base is a weak Fano toric variety, or a toric resolution of a weak Fano toric variety with blowups of height-2.

The physics of the height-2 blowups has simple interpretation as the splitting of branes. Consider a base $B$ with intersecting $I_0^*$ singularities, that admits a GWCL. Such an intersection over a curve $C$ has $MOV_C(f,g) = (4,6)$, and therefore admit a crepant base change, to produce a base $B^{'}$ without such intersections. The change of base to $B^{'}$ introduces a new divisor in $B^{'}$, which one can shrink to zero size to recover $B$, and so we should not be surprised that the resolved geometry admits a GWCL as well. However, $B^{'}$ also admits another GWCL, distinct from that of $B$, that results in non-intersecting $I_0^*$ fibers. Clearly, such a phenomena only occurs for height 2 blowups, as $B$ would not admit further base changes obtained by blowing up intersections between the exceptional divisor and the original $I_0^*$ singularities. In particular, such a $B$ must have OOV of $(6,9)$ at triple intersections $p$ of $I_0^*$ singularities which is not sufficient to admit a base change by blowing up $p$. For that to occur, the OOV must be $\geq (8,12)$;
see the Appendix of \cite{Halverson:2017ffz} for a detailed
discussion.

Let us now turn to bounding the fraction of tree toric bases that admit a GWCL. The ensemble is overwhelmingly generated by trees over $\Delta^\circ_1$ and $\Delta^\circ_2$, where a triangulation of  each of these polytopes has 72 2-simplices and 108 1-simplices. Out of the 82 possible edge trees 80 of them have a leaf of at least height 3, and of the 41,873,645 face trees 41,873,644 have a leaf of at least height 3, The fraction of geometries that admit a GWCLis  therefore calculated as
\begin{equation}
\frac{N_{\text{GWCL}}}{N_{\text{Total}}} = \left ( \frac{1}{41873645}\right)^{72} \times \left( \frac{2}{82}\right)^{108} \leq 1.1 \times 10^{-723}\,.
\end{equation}

Therefore, the fraction of geometries in the tree ensemble that allow for a GWCL make up an absolutely minuscule fraction of the total geometries.

\subsection{The Sen limit}\label{sec:senlimit}

Recall from Section~\ref{sec:weakcup} that a Sen limit is possible when all the fibers are either smooth, $I_n$, $I_0^*$, or $I^{*}_n$, which is more general than the GWCL. However, we will find that the bases that allow for a Sen limit are still very constrained. We will proceed in the same manner as before. We consider the effect of a base-changing resolution, which forces certain multiplicities of vanishing of $f$ and $g$ on divisors corresponding to points interior to faces and edges. As before, any blowup along an edge or a face forces at least type $II$ fibers on every point interior to that edge or face. Let the divisors corresponding to such interior points be $D_i$. Assume such a blowup has been done, in which case the fiber type is at least type $II$ for all $D_i$. Then each fiber above the $D_i$ must be tuned to $I_n^*$ to allow for a Sen limit. A necessary condition is then $f$ and $g$ vanish to at least orders $2$ and $3$, respectively, with $MOV (\Delta) = 6$, and so we have a similar set of conditions as in the case of the GWCL, the only difference being that we allow for an $I_1$ locus in the case of a Sen limit.

\begin{figure}[ht]
  \centering
  \begin{tikzpicture}[scale=1]
  
  \draw[line width=0.3mm] (-2,0) -- (0,0);
  \draw[line width=0.3mm] (0,0) -- (2,0);
   \draw[line width=0.3mm] (2,0) -- (4,0);
    \draw[line width=0.3mm] (4,0) -- (6,0); 
    \draw[line width=0.3mm] (6,0) -- (8,0);
     \draw[line width=0.3mm] (8,0) -- (10,0);
    \node[draw,circle,inner sep=2pt,fill,black] at (0,0) {};
    \node[draw,circle,inner sep=2pt,fill,black] at (2,0) {};
    \node[draw,circle,inner sep=2pt,fill,black] at (4,0) {};
    \node[draw,circle,inner sep=2pt,fill,black] at (6,0) {};
    \node[draw,circle,inner sep=2pt,fill,black] at (8,0) {};

   \node [label={[xshift=-0cm, yshift=-1cm]$D_{n-2}$}] at (0,0) {};
     \node [label={[xshift=-0cm, yshift=-1cm]$D_{n-1}$}] at (2,0) {};
     \node [label={[xshift=-0cm, yshift=-1cm]$D_{n}$}] at (4,0) {};
     \node [label={[xshift=-0cm, yshift=-1cm]$D_{n+1}$}] at (6,0) {};
      \node [label={[xshift=-0cm, yshift=-1cm]$D_{n+2}$}] at (8,0) {};
          
      \node [label={[xshift=-0cm, yshift=-1.7cm]$-2 + 2p$}] at (0,0) {};
     \node [label={[xshift=-0cm, yshift=-1.7cm]$-2 + p$}] at (2,0) {};
     \node [label={[xshift=-0cm, yshift=-1.7cm]$-2$}] at (4,0) {};
     \node [label={[xshift=-0cm, yshift=-1.7cm]$-2 - p$}] at (6,0) {};
      \node [label={[xshift=-0cm, yshift=-1.7cm]$-2 - 2p$}] at (8,0) {};

  \end{tikzpicture}
  
  \caption{\label{fig:cont1} An edge $e$ in $\Delta^\circ$ with several interior points, with the corresponding dot products $m \cdot v_i$ labeled. For some $m$ we have $ m \cdot v_n = -2$. The dot products $ m\cdot v_i$ for the rest of the $v_i$ on $e$ are constrained by linearity. Here $p$ is an integer.}
  \end{figure}
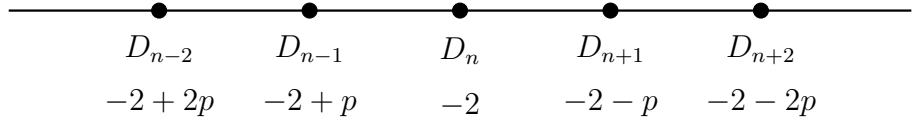

First we consider an edge $e$ of the polytope with at least one interior point. Let the points along the edge correspond to coordinates $x_i$, with corresponding divisors $D_i$. Then there must be an $m_f \in \Delta_f$, corresponding to a monomial in $f$, such that $m_f\cdot v_n = -2$, or there must be an $m_g \in \Delta_g$, corresponding to a monomial in $g$, such that $m_g\cdot v_n = -3$. Without loss of generality let us first assume the former is true, since both analyses are nearly identical and yield the same constraints. Linearity constrains the dot products between such an $m$ and the rest of the $v_i$ as displayed in Figure~\ref{fig:cont1}. 

First assume that $p > 0$ for this $m$. If we perform a height-2 blow up along a one-simplex to the right of $v_n$ then this eliminates such an $m$, as $ m\cdot ( v_n + v_{n+1})  = (-2) + (-2 - p) < -4$ for $p >0$. Similarly, if $p < 0$  then a height-2 blowup to the left of $v_n$ will eliminate such an $m$. Let us perform height-2 blowups on both the left and right of $v_n$, i.e. $p=0$. It then must be the case that $ m \cdot v_i  = -2$ for all $v_i \in e$. However, under this hypothesis, a subsequent height $\geq 3$ blowup anywhere along $e$ will eliminate any such $m$. Indeed, any height $3$ blowup leads to a ray of the form $2v_{i} + v_{i+1}$ and hence we have $ m \cdot (2v_{i} + v_{i+1})  = -6$. This violates the condition for $m$ to correspond to a monomial in $f$: $m\cdot v_a \leq -4\, \forall \, v_a \in \Delta^\circ$, and therefore the monomials protecting higher than $I_0^{*}$ vanishing along the $v_i$ are eliminated by such a blowup.

If there are at least three points interior to an edge $e$ one can say more. Let us consider a divisor $D$ corresponding to a point $p$ in the interior of $e$ that is not adjacent to either of the vertices of $e$. Assume there is a monomial, corresponding to $m_f \in \Delta_f$, that has MOV of $2$ along $D$ in $f$, which protects against higher than $I_0^{*}$ $MOV$ in $f$.  Blowing up along $e$ then forces at least type $II$ vanishing along $D$, and any other divisors corresponding to interior points of $e$. However, linearity then requires $ m, \cdot v_i = -2$ for all $v_i$ interior to $e$, which brings us to the same situation as as in a GWCL. Therefore a single blowup of height $\geq 3$ above $e$ will spoil a Sen limit.

There is a similar condition for 2-simplex blowups, where the 2-simplices are interior to a facet $F$. A 2-simplex blowup involving only points interior to a face will force $> I_n^*$ vanishing in both $f$ and $g$ on at least one divisor. 

\begin{figure}[t]
\begin{center}
\begin{tikzpicture}[scale=3]
\draw[thick,color=Black] (0,0) -- (3,0) -- (0,3) -- cycle;
\draw[thick,color=Black] (0,.5) -- (2.5,.5);
\draw[thick,color=Black] (0,1) -- (2,1);
\draw[thick,color=Black] (0,1.5) -- (1.5,1.5);
\draw[thick,color=Black] (0,2) -- (1,2);
\draw[thick,color=Black] (0,2.5) -- (.5,2.5);
\draw[thick,color=Black] (.5,0) -- (.5,2.5);
\draw[thick,color=Black] (1,0) -- (1,2);
\draw[thick,color=Black] (1.5,0) -- (1.5,1.5);
\draw[thick,color=Black] (2,0) -- (2,1);
\draw[thick,color=Black] (2.5,0) -- (2.5,.5);
\draw[thick,color=Black] (0,2) -- (.5,2.5);
\draw[thick,color=Black] (0,1) -- (1,2);
\draw[thick,color=Black] (0,0) -- (1.5,1.5);
\draw[thick,color=Black] (1,0) -- (2,1);
\draw[thick,color=Black] (2,0) -- (2.5,.5);
\draw[thick,color=Black] (0,1.5) -- (.5,2);
\draw[thick,color=Black] (0,.5) -- (1,1.5);
\draw[thick,color=Black] (.5,0) -- (1.5,1);
\draw[thick,color=Black] (1.5,0) -- (2,.5);
\fill (0,0) circle (.5mm); \fill (0,.5) circle (.5mm); \fill (0,1) circle (.5mm);
\fill (0,1.5) circle (.5mm); \fill (0,2) circle (.5mm); \fill (0,2.5) circle (.5mm);
\fill (0,3) circle (.5mm);
\fill (.5,0) circle (.5mm); \fill (.5,.5) circle (.5mm); \fill (.5,1) circle (.5mm);
\fill (.5,1.5) circle (.5mm); \fill (.5,2) circle (.5mm); \fill (.5,2.5) circle (.5mm);
\fill (1,0) circle (.5mm); \fill (1,.5) circle (.5mm); \fill (1,1) circle (.5mm);
\fill (1,1.5) circle (.5mm); \fill (1,2) circle (.5mm); 
\fill (1.5,0) circle (.5mm); \fill (1.5,.5) circle (.5mm); \fill (1.5,1) circle (.5mm);
\fill (1.5,1.5) circle (.5mm); 
\fill (2,0) circle (.5mm); \fill (2,.5) circle (.5mm); \fill (2,1) circle (.5mm);
\fill (2.5,0) circle (.5mm); \fill (2.5,.5) circle (.5mm);
\fill (3,0) circle (.5mm);
\node [label={[xshift=-.32cm, yshift=.05cm]$v_{0}$}] at (1,1) {};
\end{tikzpicture}

\caption{The largest facet $F_1$ of the 3d reflexive polytope $\Delta_1^\circ$, with an arbitrary triangulation. A type $II$ fiber above the divisor corresponding to $v_0$ places strong constraints on the existence of the Sen limit.}
\label{fig:bigfacetbigone1}
\end{center}
\end{figure}
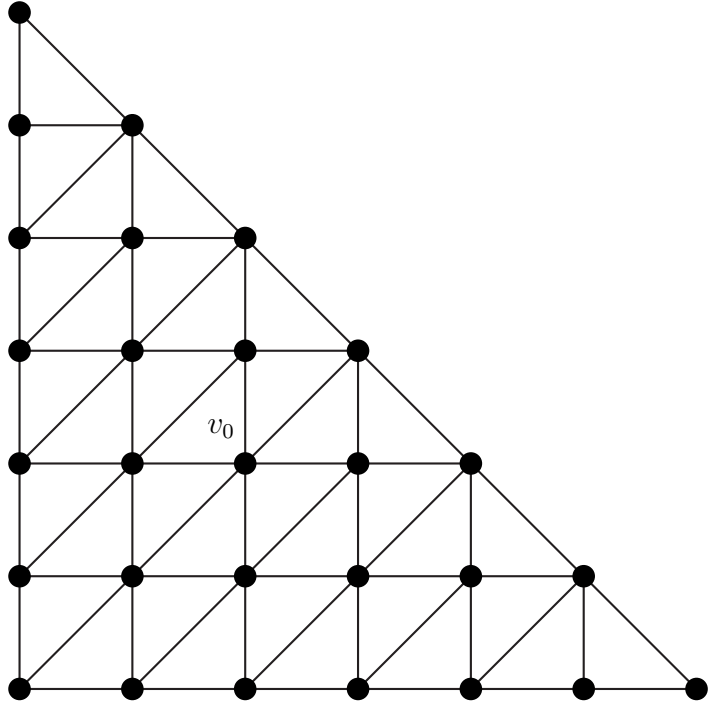

An even stronger condition occurs nearly universally in the tree ensemble.  First consider trees built over $\Delta^\circ_1$. The largest facet $F_1$ of  $\Delta^\circ_1$ is shown in Figure~\ref{fig:bigfacetbigone1}. 
After a height $\geq 2$ blowup anywhere on the face all of the points interior to $F_1$ will have at least type $II$ fibers above them. Therefore, to admit a Sen limit, we must tune to at least $I_0^*$ on all of these points. The point in the middle $v_0$, corresponding to a divisor $D_0$, has neighbors in both the vertical and horizontal directions with $I_0^*$ fibers. This implies that there is at least one monomial in $f$ that vanishes to multiplicity $2$ along $D_0$, or at least one monomial in $g$ that vanishes to multiplicity $3$ along $D_0$. Without loss of generality let us assume the former is true. We then have $m \cdot v_0 = -2$ for some $m$. However, the fact that we have all $I_0^*$ on the points $v_i$ interior to $F_1$ implies $m \cdot v_i  = -2$, and therefore $ m \cdot F_1  = -2$. It is clear then any height $\geq 3$ blowup on $F_1$ will eliminate this monomial, and force greater than $I_0^*$ on the divisor $D_0$. Therefore a height $\geq 3$ blowup on $F_1$ will obstruct the Sen limit. 

\begin{figure}[t]
\begin{center}
\begin{tikzpicture}[scale=2]
\draw[thick,color=Black] (1,3) -- (7,3) -- (7,0) -- cycle;
\draw[thick,color=Black] (3,2) -- (7,2);
\draw[thick,color=Black] (5,1) -- (7,1);
\draw[thick,color=Black] (6.5,1) -- (6.5,3);
\draw[thick,color=Black] (6,1) -- (6,3);
\draw[thick,color=Black] (5.5,1) -- (5.5,3);
\draw[thick,color=Black] (5,1) -- (5,3);
\draw[thick,color=Black] (4.5,2) -- (4.5,3);
\draw[thick,color=Black] (4,2) -- (4,3);
\draw[thick,color=Black] (3.5,2) -- (3.5,3);
\draw[thick,color=Black] (3,2) -- (3,3);
\draw[thick,color=Black] (7,1) -- (6,3);
\draw[thick,color=Black] (6.5,1) -- (5.5,3);
\draw[thick,color=Black] (6,1) -- (5,3);
\draw[thick,color=Black] (5.5,1) -- (4.5,3);
\draw[thick,color=Black] (5,1) -- (4,3);
\draw[thick,color=Black] (7,2) -- (6.5,3);
\draw[thick,color=Black] (4,2) -- (3.5,3);
\draw[thick,color=Black] (3.5,2) -- (3,3);
\draw[thick,color=Black] (7,0) -- (6.5,1);
\draw[thick,color=Black] (7,0) -- (6,1);
\draw[thick,color=Black] (7,0) -- (5.5,1);
\draw[thick,color=Black] (5,1) -- (4.5,2);
\draw[thick,color=Black] (5,1) -- (4,2);
\draw[thick,color=Black] (5,1) -- (3.5,2);
\draw[thick,color=Black] (3,2) -- (2.5,3);
\draw[thick,color=Black] (3,2) -- (2,3);
\draw[thick,color=Black] (3,2) -- (1.5,3);
\fill (1,3) circle (.5mm); \fill (1.5,3) circle (.5mm); \fill (2,3) circle (.5mm); \fill (2.5,3) circle (.5mm);
\fill (3,3) circle (.5mm); \fill (3.5,3) circle (.5mm); \fill (4,3) circle (.5mm); \fill (4.5,3) circle (.5mm);
\fill (5,3) circle (.5mm); \fill (5.5,3) circle (.5mm); \fill (6,3) circle (.5mm); \fill (6.5,3) circle (.5mm);
\fill (7,3) circle (.5mm);
\fill (7,2) circle (.5mm);
\fill (7,1) circle (.5mm);
\fill (7,0) circle (.5mm);
\fill (6.5,1) circle (.5mm);
\fill (6,1) circle (.5mm);
\fill (5.5,1) circle (.5mm);
\fill (5,1) circle (.5mm);
\fill (6.5,2) circle (.5mm);
\fill (6,2) circle (.5mm);
\fill (5.5,2) circle (.5mm);
\fill (5,2) circle (.5mm);
\fill (4.5,2) circle (.5mm);
\fill (4,2) circle (.5mm);
\fill (3.5,2) circle (.5mm);
\fill (3,2) circle (.5mm);
\node [label={[xshift=-.5cm, yshift=.0cm]$v_{0}$}] at (5,2) {};
\node [label={[xshift=-.0cm, yshift=.0cm]$\langle m, \cdot \rangle = -2$}] at (8,1.8) {};
\node [label={[xshift=-.0cm, yshift=.0cm]$\langle m, \cdot \rangle = -3$}] at (8,.8) {};
\node [label={[xshift=-.0cm, yshift=.0cm]$\langle m, \cdot \rangle = -4$}] at (8,-.2) {};
\node [label={[xshift=-.0cm, yshift=.0cm]$\langle m, \cdot \rangle = -1$}] at (8,2.8) {};

\end{tikzpicture}

\caption{The largest facet $F_2$ of the 3d reflexive polytope $\Delta_1^\circ$, with an arbitrary triangulation. The dot products of the special monomial in $\Delta_f$ with the points along each horizontal line are shown to the right of $F_2$.}\label{fig:bigfacetbigone2}
\end{center}
\end{figure}
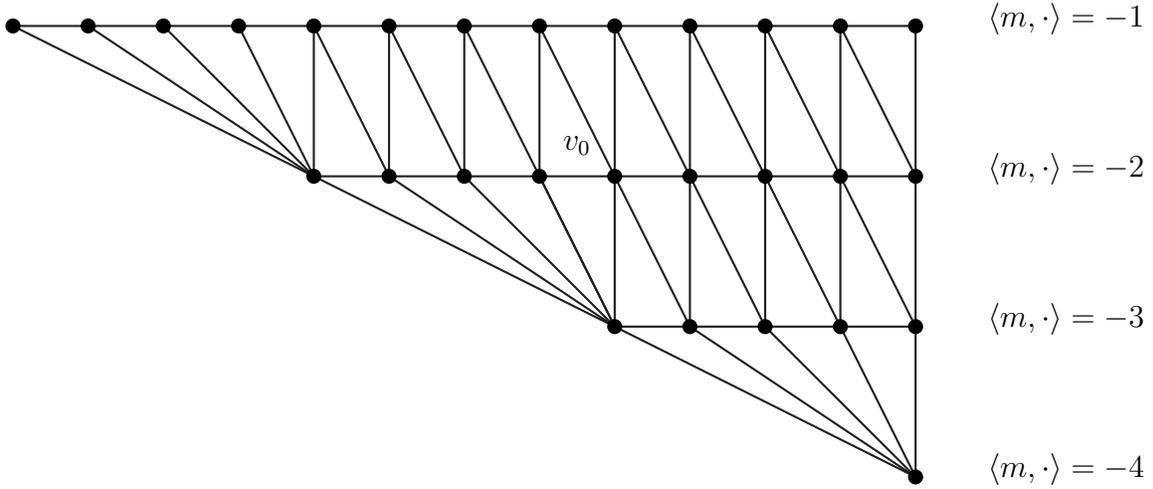

A similar analysis can be done for $\Delta^\circ_2$. The largest facet $F_2$ of  $\Delta^\circ_2$ is shown in Figure~\ref{fig:bigfacetbigone2}. However, there is no interior point that is bounded both horizontally and vertically by interior points. Instead we consider a point in the second from the top horizontal row of points $h_2$, that is bounded horizontally by interior points. An example of such a point is labelled $v_0$  in Figure~\ref{fig:bigfacetbigone2}. Taking the same approach as above, for there to exist a Sen limit there must be a monomial in $f$, corresponding to a vector $m$, such that $ m\cdot v_0  = -2$. To have $I_0^*$ on each point interior to $F_2$ then implies $ m\cdot h_2  = -2$, and furthermore $m\cdot h_3 = -3$, where $h_3$ is the row of horizontal points third from the top, and so forth. The corresponding dot products are displayed on the right in Figure~\ref{fig:bigfacetbigone2}. This is fairly constraining, but one can make the same argument for a point in $h_3$, which switches the ordering of dot products vertically. Then one can immediately conclude that a height $\geq 3$ blowup on $F_2$ will obstruct a Sen limit. 

Therefore, we are able to provide a rather strict upper bound on the probability of finding a base that admits a Sen limit in our ensemble, similar to what we found in the GWCL case. Since either a single height-3 blowup internal to one of the large facets $F_1$ or $F_2$ eliminates the possibility of a Sen limit,  an upper bound on the a fraction of tree geometries admit a weakly coupled Sen limit is calculated to be:
\begin{equation}
\frac{N_{\text{Sen}}}{N_{\text{Total}}} = \left ( \frac{1}{41873645}\right)^{36} \times   \left ( \frac{2}{82}\right)^{63} \leq 1.0 \times10^{-376}\,.
\end{equation}

We can strengthen this slightly by using the additional constraints that the other facets provide. By requiring that there are height-2 blowups along two 1-simplices on the same edge, coupled with a height-3 blowup along that same edge, on the three edges away from $F_1$ and $F_2$, one picks up an additional factor of $3\times10^{-15}$, yielding a fraction of $3\times10^{-391}$. There are other constraints that significantly reduce this number further, but these are triangulation dependent.

\vspace{.5cm}
Let us summarize the results. The Sen limit is spoiled in general for
\begin{itemize}
\item A height-2 blowup along two 1-simplices on the same edge, coupled with a height-3 blowup along that same edge.
\item A height-3 blowup along a 2-simplex strictly interior to a face.
\end{itemize}
The conditions in the tree ensemble are even stricter, as we can simply consider the largest facets $F_1$ and $F_2$ of the largest 3d reflexive polytopes $\Delta^\circ_1$ and $\Delta^\circ_2$. For the tree ensemble, the Sen limit is spoiled for any height-3 blowup along $F_1$ or $F_2$.  

\section{ Weak Coupling Limits on More General Algebraic Bases}\label{sec:alg}

In the previous section we considered toric threefold bases, as the combinatorial properties of toric varieties allowed us to greatly simplify the computation for determining whether a base admits either of the weak coupling limits. One of the key structures of toric varieties that allowed us to determine this was that the smooth toric varieties could be interpreted as crepant resolutions of a singular, more minimal toric variety. The resolution then introduced rigid exceptional divisors, and further base-changing resolutions forced NH7s on these exceptional divisors. 
The toric case can be viewed as a two-step process. First, we start with a singular base $B^{'}$, and crepantly resolve it to a smooth base $B$, without introducing any NH7. 

We then build trees of geometries on top of $B$ by performing base-changing resolution. We follow the same procedure in the non-toric case. We will start with a possibly singular $B^{'}$, and resolve it to a new space $B$. Our assumption, without loss of generality, will be that resolving to $B$ does not produce any non-Higgsable 7-branes, so that the generic Calabi-Yau elliptic fibration over $B$ is smooth. In performing these blowups we naturally generalize the notion of the ``height'' of a divisor, which we can define inductively. Let a divisor be a blowup along the intersection $n$ divisors. The height of the blowup divisor is given by the sum of the $n$ heights of the divisors whose intersection defines the locus to be blown up. The induction terminates by defining the divisors of a \textit{ground geometry} to have height-1. Here a geometry is a ground geometry if the generic elliptic fibration over it has no non-Higgsable 7-branes.

The structure of $B$ will be partially determined by the fact that $B$ is a crepant resolution of $B^{'}$, which will allow us to analyze whether blowups of $B$ allow for weak coupling limits. This itself is quite convenient: while a singular variety may admit many crepant resolutions, some of the features of the resolved spaces can be read off from the singular space itself, without needing to explicitly resolve. The toric examples in the previous section are such an example; the sections of $\mathcal{O}(-nK_B)$ are calculated only using the data of $\Delta^\circ$, and the existence of weak-coupling limits did not strongly depend on the resolution, i.e. on the choice of a FRST. We wish to proceed in a similar manner, in the spirit of minimal models, with base geometries that are non-toric.

We will generalize this procedure by constructing global geometries via gluing together local patches. In some cases this will allow us to use the tools of toric geometry locally, without requiring the global, compact variety to be toric. We will consider local patches of the form $\mathbb{C}^3 / G$, where $G$ is a finite subgroup of $SL(3, \mathbb{C})$. Local patches will then be glued together to form a global K\"ahler manifold $X$. This type of orbifold is especially convenient to work with because it is known to admit a crepant resolution~\cite{ROAN1996489}. In addition, if $G$ is abelian then $\mathbb{C}^3 / G$ can be resolved using toric methods~\cite{roan1989}.  Let $X$ be such an orbifold. The singular points of $X$ can be divided into two categories~\cite{Joyce2003}:
\begin{enumerate}
\item Singular points modeled on $(\mathbb{C}^2 / H) \times \mathbb{C}$, where $H$ is a finite subgroup of $SU(2)$.
\item Singular points not of the first type.
\end{enumerate}
The singular points of the second type are actually discrete isolated points, which implies that $G$ is abelian~\cite{Degeratu_flopsof}, and therefore locally admit a toric description. Singularities of the first type correspond to $ADE$ singularities fibered over a curve. Only the $A$-type singularities are realizable as toric.

Since we do not have a global description of $X$ we do not have a way to determine global sections of the anticanonical bundle and its powers. Instead, we will assume the most general local form and understand the behavior of sections as we resolve. This is sufficient because the general form can only be further restricted by gluing the local patch into a compact variety.

\subsection{Isolated singularities}
We begin by discussing the isolated rational points, which admit a toric description. The space is locally Calabi-Yau, and since we will be considering only crepant resolutions all rays, including those corresponding to blowups, can be taken to lie on a 2d plane, which we will informally refer to as a facet $F$. The singular point is located at $x_1 = x_2 = x_3 = 0$, which corresponds to $3d$ cone $C$ formed by the rays $v_1$, $v_2$, $v_3$. 

As an illustrative example we first consider the case where there is a single interior point $v_z$ to $F$, such that $v_z = (v_{x_1} +v_{x_2} + v_{x_3})/3$. Resolution of $C$ then involves introducing the ray $v_z$, and subdividing $F$. Let us take a local section of $\mathcal{O}(-4 K)$ of the form\footnote{In this section we suppress all factors that do not include the local coordinates, as they will not affect the calculations.} $x_1^a x_2^b x_3^c$. The toric resolution promotes this to $x_1^a x_2^b x_3^c z^{(a+b+c)/3}$. If we tune an $I_0^*$ fiber on $D_z$, such that $f$ vanishes to order 2 along $z = 0$, then we find $a+ b+ c = 6$, which with the assumption $I_0^{*}$ or smooth fibers along the other divisors, then implies $a = b = c =2$. There is an analogous story for the monomials in $g$. Therefore, we find that introducing an $I_0^*$ fiber on $D_z$ forces an $I_0^*$ fiber on $D_{x_1}$, $D_{x_2}$, and $D_{x_3}$. 

This generalizes readily to more general singularities. Consider a facet $F$ with many interior points. All points in $F$ can be written as $v_{z_i} = \alpha_i v_{x_1} + \beta_i v_{x_2} + (1 - \alpha_i -\beta_i) v_{x_3}$ for non-negative $\alpha_i, \beta_i$. Under a resolution, a section $x_1^a x_2^b x_3^c$ is promoted to 
\begin{equation}\label{eqn:pointres}
x_1^a x_2^b x_3^c \rightarrow x_1^a x_2^b x_3^c z_1^{(a \alpha_1 + b \beta_1 + c(1- \alpha_1 - \beta_1))}\dots z_n^{(a \alpha_n + b \beta_n + c(1- \alpha_n - \beta_n))}\, 
\end{equation}
where $z_1, \dots, z_n$ are the projective coordinates introduced in the resolution. Let us tune an $I_0^*$ on $D_{z_1}$. We then have $a \alpha_1 + b \beta_1 + c(1- \alpha_1 - \beta_1) = 2$. In order to have a GWCL we need $a,b,c \in \{0, 2\}$, and we see that $a \alpha_1 + b \beta_1 + c(1- \alpha_1 - \beta_1) = 2$ then requires $a = b = c= 2$, which implies the section is of the form
\begin{equation}
x_1^2 x_2^2 x_3^2 z_1^2\dots z_n^2\, 
\end{equation}
Therefore, the existence of an $I_0^*$ on any of the $D_{z_i}$, along with requiring a GWCL, forces an $I_0^*$ on all divisors corresponding to points on $F$.

We now study the existence of a GWCL  after building trees above such a patch. First let us blow up a point. Blowing up the intersection of any three of the $D_{z_i}$ requires us to tune $f$ to order 8 along the intersection. From Eq.~\ref{eqn:pointres} we can see by demanding any order of vanishing along an intersection of the $D_{z_i}$ this clearly forces at least a type II fiber on all the $D_{z_i}$, as the powers of the $z_i$ in Eq.~\ref{eqn:pointres} all become non-zero. We must therefore tune an $I_0^*$ on all of $D_{z_i}$ to get a GWCL. This forces an $I_0^*$ on $D_{x_1}$, $D_{x_2}$, and $D_{x_3}$ as well, as requiring the absence of an $I_1$ locus forces $a = b = c= 2$. Thus, by
contradiction we see that a GWCL is not compatible with
such a blowup. 
However, we immediately see the assumption of only $I_0^{*}$ on each of these divisors is incompatible with the required OOV of $(8,12)$ to crepantly blowup the point, and so we see that this height-3 blowup spoils the GWCL.

The Sen limit case is most constrained for isolated singularities that require at least three exceptional divisors to crepantly resolve, similar to the toric case. A height $\geq 2$ blowup on $F$ will force at least $II$ vanishing on all of the exceptional divisors, and we therefore need to tune to $I_0^{*}$ vanishing on such divisors. In this case tuning to $I_0^{*}$ vanishing on all the exceptional divisors forces an $I_0^*$ above $D_{x_1}$, $D_{x_2}$, and $D_{x_3}$. One can see this by noting that tuning an $I_0^{*}$ on three exceptional divisors is done by solving the equations, for instance, for vanishing to multiplicity two in $f$:
\begin{align}\label{eqn:three}
& a \alpha_1 + b \beta_1 + c(1- \alpha_1 - \beta_1) = 2\, \nonumber \\
& a \alpha_2 + b \beta_2 + c(1- \alpha_2 - \beta_2) = 2\, \nonumber \\
& a \alpha_3 + b \beta_3 + c(1- \alpha_3 - \beta_3) = 2\, 
\end{align}
The unique solution is $a = b = c=  2$, and so any height-3 blowup on $F$, including above a curve, spoils a Sen limit, as one must then turn off all the monomials with $a = b = c =2$ to obtain the necessary multiplicity eight vanishing in $f$, thereby tuning to $> I_0^{*}$ vanishing in $f$. A similar argument applies for monomials in $g$. This confirms the results for the $\Delta^\circ_i$ that we found at the end of Section~\ref{sec:senlimit}, for which the large facets $F_i$ are examples of resolutions of isolated singularities, from a more general framework.

\subsection{$A_n$ over a curve}
\begin{figure}[ht]
  \centering
  \begin{tikzpicture}[scale=1]
  
   \draw[line width=0.3mm] (2,0) -- (4,0);
     \draw[line width=0.3mm] (2,0) -- (3,1);
        \draw[line width=0.3mm] (3,1) -- (4,0);
    \draw[line width=0.3mm] (4,0) -- (6,0); 
      \draw[line width=0.3mm] (6,0) -- (8,0); 
    \node[draw,circle,inner sep=2pt,fill,black] at (2,0) {};
    \node[draw,circle,inner sep=2pt,fill,black] at (4,0) {};
    \node[draw,circle,inner sep=2pt,fill,black] at (6,0) {};
    \node[draw,circle,inner sep=2pt,fill,black] at (3,1) {};

     \node [label={[xshift=-0cm, yshift=-1cm]$x_1$}] at (2,0) {};
     \node [label={[xshift=-0cm, yshift=-1cm]$z_1$}] at (4,0) {};
     \node [label={[xshift=-0cm, yshift=-1cm]$z_2$}] at (6,0) {};
     \node [label={[xshift=-0cm, yshift=-.8cm]$e$}] at (3,1) {};
       \end{tikzpicture}
  
  \caption{\label{fig:bu1} A blowup of the curve $x_1= z_1 = 0$, which introduces the exceptional coordinate $e$. This forces at least type $II$ fibers above all of the $D_{z_i}$.}
  \end{figure}
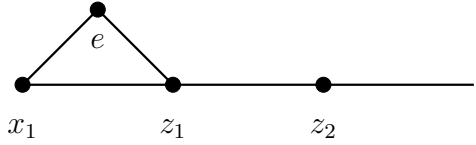

Let us now turn to the singularities of type (1). We begin with $A$-type singularities, as these admit toric resolutions locally.  The toric description of an $A_n$ singularity is well-known; the fan has rays $(1,0)$ and $(1,n+1)$, with corresponding coordinates $x_1$ and $x_2$,  and the resolution introduces the rays $(1,1), \dots, (1,n)$, with corresponding exceptional coordinates $z_i$.  Let a curve $C$ be the intersection of $D_{z_1}$ with another divisor, which could be $D_{x_1}$, $D_{x_2}$, or another exceptional divisor. In order for the base change to be crepant we require that the $MOV_C (f,g) \geq (4,6)$. Let us focus on monomials in $f$, which are all of the form
\begin{equation}\label{eqn:sing}
x_1^a x_2^b z_1^{\alpha_1a + (1 -\alpha_1)b}\dots z_n^{\alpha_n a+ (1 -\alpha_n)b}\, 
\end{equation}
for positive $\alpha_i$. First, let us tune a $(4,6)$ curve $C$ on $x_1 = z_1 = 0$ by tuning toric monomials in $f$ and $g$, and then perform a base changing resolution over $C$ by blowing up the locus $x_1=z_1=0$, which is depicted in Fig~\ref{fig:bu1}. Note that $0 < \alpha_i < 1$, and so we see that the tuning required for a crepant resolution forces at least a type $II$ fiber on $D_{z_1}$, because the mentioned tuning forces $a\neq 0$ or $b\neq 0$. In this case, performing the crepant base change promotes Eq.~\ref{eqn:sing} to
\begin{equation}
  x_1^a  x_2^b z_1^{\alpha_1 a + (1-\alpha_1)b} e^{(1+\alpha_1)a + (1-\alpha_1)b - 4}\dots z_n^{\alpha_na+ (1 -\alpha_n)b}\, 
\end{equation}
where the $-4$ arises from the crepant base changing resolution\footnote{See \cite{Halverson:2017ffz} for further details.}.

In a similar way, we could also tune a $(4,6)$ fiber above the curve $z_1  = z_2 = 0$ and perform a base changing resolution, as depicted in Fig.~\ref{fig:bu1}.
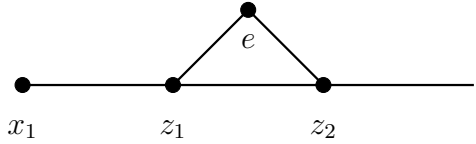
\begin{figure}[ht]
  \centering
  \begin{tikzpicture}[scale=1]
  
   \draw[line width=0.3mm] (2,0) -- (4,0);
     \draw[line width=0.3mm] (4,0) -- (5,1);
        \draw[line width=0.3mm] (5,1) -- (6,0);
    \draw[line width=0.3mm] (4,0) -- (6,0); 
      \draw[line width=0.3mm] (6,0) -- (8,0); 
    \node[draw,circle,inner sep=2pt,fill,black] at (2,0) {};
    \node[draw,circle,inner sep=2pt,fill,black] at (4,0) {};
    \node[draw,circle,inner sep=2pt,fill,black] at (6,0) {};
    \node[draw,circle,inner sep=2pt,fill,black] at (5,1) {};

     \node [label={[xshift=-0cm, yshift=-1cm]$x_1$}] at (2,0) {};
     \node [label={[xshift=-0cm, yshift=-1cm]$z_1$}] at (4,0) {};
     \node [label={[xshift=-0cm, yshift=-1cm]$z_2$}] at (6,0) {};
     \node [label={[xshift=-0cm, yshift=-.8cm]$e$}] at (5,1) {};
       \end{tikzpicture}
  
  \caption{\label{fig:bu2} A blowup of the curve $z_1 = z_2 =  0$, which introduces the exceptional coordinate $e$. This forces at least type $II$ fibers above all of the $D_{z_i}$.}
  \end{figure}
 This also results in type $II$ fibers above all of the $D_{z_i}$, and so we see that blowing any curve $C$ that is the intersection of a $z_i$ with another divisor forces at least a type $II$ fiber on all of the resolution divisors $D_{z_i}$. In this case, performing the crepant base change promotes Eq.~\ref{eqn:sing} to
\begin{equation}
x_1^a  x_2^b z_1^{\alpha_1 a + (1-\alpha_1)b} z_2^{\alpha_2 a + (1-\alpha_2)b} e^{(\alpha_1 + \alpha_2)a+ (2-\alpha_1 - \alpha_2)b - 4}\dots z_n^{\alpha_n a + (1 -\alpha_n)b}\, 
\end{equation}

Having performed either resolution, we can now discuss weak coupling limits. We first discuss the GWCL.  As the resolution forces a type $II$ fiber over all the $D_{z_i}$ we then require an $I_0^*$ fiber above $D_{z_1}, \dots, D_{z_n}$. Requiring only smooth or $I_0^{*}$ fibers then forces an $I_0^{*}$ above $D_{x_1}$ and $D_{x_2}$, and so any monomial in $f$ takes the form
\begin{equation}\label{eqn:resolved}
x_1^2 x_2^2 z_1^2\dots z_n^2\, 
\end{equation}
Note that by demanding a GWCL the exceptional coordinate $e$ arising from the resolution does not appear in Eq.~\ref{eqn:resolved} due to the $-4$, since $a=b=2$. If we then wish to perform a height-3 blowup along $D_e \cap D_i$, for some divisor $D_i$, we would need to tune a $(4,6)$ curve along $D_e \cap D_i$.  This would clearly spoil the form of Eq.~\ref{eqn:resolved},
as those monomials would no longer appear in the Weierstrass equation, and therefore a height-3 blowup over any such $C$ spoils the GWCL.

In the case of a Sen limit then we demand an $I_0^*$ or $I_n^{*}$ fiber above $D_{z_1}, \dots, D_{z_n}$. Let us assume that there is a monomial in $f$ preventing $MOV_{z_a}(f) >2$. The powers of the other variables are constrained by linearity, in the exact same way as the toric edge case, demonstrated in Fig.~\ref{fig:cont1}. Such a monomial takes the form
\begin{equation}
\dots z_{a-2}^{2-2p}z_{a-1}^{2-p}z_{a}^{2}z_{a+1}^{2+p}z_{a+2}^{2+2p}\dots \, 
\end{equation}
for integer $p$, which may be positive or negative. Blowing up along two different curves that arise from the resolution of the $A_n$ singularity forces $p = 0$, in order to satisfy the assumption of $I_0^{*}$ on all the $D_{z_i}$. The argument for this is identical to the one made below Fig.~\ref{fig:cont1}. A height $\geq 3$ blowup will then eliminate this monomial, eliminating the possibility of the Sen limit. A similar argument holds for a monomial in $g$ preventing MOV of $4$ in $g$.

Remarkably, using only local models we have reproduced the results of Section~\ref{sec:toric} on weak coupling limits on toric bases, in the case of isolated singularities and $A_n$ singularities fibered over curves. This should come as no surprise, as both of these cases admit toric descriptions locally. However, the local model approach has allowed us understand the effects of the base-changing resolutions on the sections of line bundles from the point of view of the Cox ring, or the homogenous coordinate ring, as opposed to the toric-specific structure of fans and polytopes. With this in hand we can now approach the more general case of $D_n$ and $E_n$ singularities fibered over curves.

\subsection{$D_n$ and $E_n$ over a curve} 
We are finally left with singularities of type (1) for the $D$ and $E$ series. We will find that a single blowup removes the possibility of either weak coupling limit. The $D$ and $E$ series singularities, as well as their resolutions, are most conveniently expressed as complete intersections in affine space.
We first consider the case of a $D_n$ singularity, that can be realized as a hypersurface in $\mathbb{A}^3$ embedded as
\begin{equation}
X^2 + Z Y^2 + Z^{n-1} = 0\, 
\end{equation}
When $n = 2k$, $k \geq 2$, the Cox ring  is generated by~\cite{2015arXiv150201040F}:

\begin{align}\label{eqn:dn}
&Z_1 = x_1^2 z_0^{2k-2} z_1^k z_2^{k-1} z_3^{2k-3} z_4^{2k-4}\dots z_{2k-1}\,  \\ \nonumber
&Z_2 = x_2^2 z_0^{2k-2} z_1^{k-1} z_2^{k} z_3^{2k-3} z_4^{2k-4}\dots z_{2k-1}\,  \\ \nonumber
&Z_3 = x_{2k-1}^2 z_0^{2} z_1 z_2 z_3^{2} z_4^{2}\dots z_{2k-1}^2\,  \\ \nonumber
&W = x_{1} x_2 x_{2k-1} z_0^{2k-1} z_1^k z_2^k z_3^{2k-2} z_4^{2k-3}\dots z_{2k-1}^2\, 
\end{align}
Here the $z_i$ correspond to the exceptional coordinates that arise in the resolution. Let us consider a general monomial $Z_1^a Z_2 ^b Z_3^c W^d$, as a section of $\mathcal{O}(-4K)$. 
From the form of Eq.~\ref{eqn:dn} one sees that $z_0$, $z_1$, $z_2$, and $z_3$ appears in any monomial with at least power $1$, and so any crepant resolution of a $D_{2k}$ singularity already has non-Higgsable 7-branes in the general Weierstrass model over that patch, and therefore $D_{z_0}, \dots, D_{z_3}$ have at least type II fibers above them.  We therefore need to tune $I_0^{*}$ fibers above these divisors in order to realize either weak coupling limit. The multiplicities of each coordinate in a given monomial are:
\begin{align}
& mult(z_0) = (2 k - 2) a + (2 k - 2) b + 2 c + (2 k - 1) d \,  \nonumber \\
& mult(z_1) =  k a + (k - 1) b + c + k d\,  \nonumber \\
& mult(z_2) = (k - 1) a + k b + c + k d\,  \nonumber \\
& mult(z_3) =  (2 k - 3) a + (2 k - 3) b + 2 c + (2 k - 2) d\, 
\end{align}
In particular, note that $mult(z_0) - mult(z_3) = a + b + d$. Therefore, in order to tune an $I_0^{*}$ fiber above both $z_0$ and $z_3$, there must be a monomial in $f$ that is multiplicity 2 in $z_0$, and multiplicity $\geq 2$ in $z_3$. This monomial must then satisfy $a =b =d =0$, and $c = 1$, as one can see from the form of $Z_3$. However, this monomial would result in a type $II$ fiber above $z_1$ and $z_2$, which contradicts the assumptions of $I_0^{*}$ above all of the exceptional divisors. Therefore crepant resolutions of $D_{2k}$ singularities do not admit either weak coupling limit. The same result follows in a similar manner for the case of odd $n = 2k +1$. 

The absence of weak coupling limits for $E_6, E_7$, and $E_8$ singularities follows in the same way. For each case the generators of the Cox ring are presented in Table~\ref{tab:cox}. Let us consider the $E_6$ case as an example. From Table~\ref{tab:cox} we can read off that in order to prevent a greater than $I_0^{*}$ vanishing along $z_0$ there must be a monomial in $f$ of the form $r Z_1 \subset f$, where $r \in \mathbb{C}$. However, this would imply a type $II$ fiber above $z_3$ and $z_5$, spoiling both weak coupling limits. The same conclusion follows trivially for $E_7$ and $E_8$, from the multiplicity of vanishing of the $z_0$ coordinate. 

\begin{table}[t]
\begin{center}
\scalebox{1}{\begin{tabular}{|c|p{55mm}|}
\hline
$n$ & $\text{Generators}$  \\  \hline
$6$&$Z_1 = z_0^2 z_1^2z_2^2 z_3 z_4^2 z_5 x_1$ \newline   $Z_2 = z_0^4 z_1^2z_2^3 z_3^2 z_4^3 z_5^2 x_3 x_5$ \newline $ Z_3 = z_0^6 z_1^3z_2^4 z_3^2 z_4^5 z_5^4 x_3^3$ \newline $ Z_4 = z_0^6 z_1^3z_2^5 z_3^4 z_4^4 z_5^2 x_5^3$ \\ \hline 
$7$& $Z_1 = z_0^4 z_1^2z_2^3 z_3^2 z_4^3 z_5^2 z_6 x_3$ \newline   $Z_2 = z_0^{12} z_1^7z_2^8 z_3^4 z_4^9 z_5^6 z_6^3 x_1^2$ \newline $ Z_3 = z_0^9 z_1^5z_2^6 z_3^3 z_4^7 z_5^5 z_6^3 x_1 x_6$ \newline $ Z_4 = z_0^6 z_1^3z_2^4 z_3^2 z_4^5 z_5^4 z_6^3 x_6^2$ \\ \hline
$8$& $Z_1 = z_0^{15} z_1^8z_2^{10} z_3^5 z_4^{12} z_5^9 z_6^6 z_7^3 x_1$ \newline   $Z_2 = z_0^6 z_1^3z_2^4 z_3^2 z_4^5 z_5^4 z_6^3 z_7^2 x_7$ \newline $ Z_3 = z_0^{10} z_1^5z_2^7 z_3^4 z_4^8 z_5^6 z_6^4 z_7^2 x_3$  \\ \hline 
\end{tabular}}
\caption{Generators of the Cox ring for the resolutions of the $E_n$ singularities.}
\label{tab:cox}
\end{center}
\end{table}

\section{ Discussion}\label{sec:dis}

In this paper we determined conditions on the base space $B$
of an F-theory geometry that are sufficient to preclude
the existence of a Sen limit, or of a GWCL.
In the toric case, the conditions sufficient for the
absence of the limits are
\begin{itemize}
\item GWCL: any height-3 blowup on a face or an edge with at least one interior point.
\item Sen: a height-3 blowup of a point above three exceptional divisors, or a height-2 blowup along two different curves represented by 1-simplices on the same edge, coupled with a height-3 blowup on that edge.
\end{itemize}
In the tree ensemble the conditions for the Sen limit were also quite restrictive, as a height-3 blowup along either of the large facets is enough to spoil the existence of a Sen limit. Our strong tree ensemble constraints could be further strengthened by a more detailed study of the facets. 

By understanding the geometry behind these conditions in the
toric case, we were also able to move beyond the toric case to more general bases. The bases we considered are generalizations of the toric case, where local patches are constructed via crepant resolutions of orbifold singularities and then glued together. In the cases of a) isolated singularities and b) $A_n$ singularities fibered over curves,  the conditions sufficient for the absence of the weak coupling limits are the same as the toric case. This was expected, as these types of singularities admit a toric description and resolution. In the case of $D_n$ and $E_n$ singularities fibered over curves, we found that there ware no weak coupling limits, due to the behavior of the Cox ring under the resolution.

We then performed
a geometric analysis of a class of $\frac{4}{3}\times 2.96\times 10^{755}$ bases that are built from crepant base-changing resolutions of
elliptic fibrations over weak Fano toric threefolds; i.e. a set of $10^{755}$ elliptic fourfolds that are related
by topological transitions. Using the conditions we derived,
we showed that nearly all of these geometries do not admit
weak coupling limits. Specifically, the fraction admitting
a GWCL or a Sen limit are bounded above by
\begin{equation}
  \frac{N_{\text{GWCL}}}{N_{\text{Total}}} \leq 1.1 \times 10^{-723}\ \qquad \qquad 
\frac{N_{\text{Sen}}}{N_{\text{Total}}}  \leq  3.0  \times 10^{-391}\,.
\end{equation}
This strengthens the previous weak coupling result of \cite{Halverson:2017ffz} that was based on non-Higgsable clusters.
While that work showed that generic points in
the complex structure moduli of $X$ over generic bases
are strongly coupled, this work showed that essentially none
of them admit weak coupling limits; i.e. not only are the generic
points strongly coupled, but furthermore there are no subloci in complex
structure that become weakly coupled. This was not forbidden
a priori, as the bases that give rise to NHC in principle could
all be enhanced to $I_n^*$ fibers and admit a Sen limit or GWCL
limit.

It is likely that Sen limits or GWCLs are similarly
rare in the complete set of bases for four-dimensional F-theory
compactifications, which is currently unknown. The reasoning
is that the prevalence of NHC is correlated strongly with
moving away from Fano or weak-Fano threefolds via topological
transitions. Once a cluster exists, the existence of a Sen limit
or GWCL requires that the singular fibers that exist
for generic moduli are less singular
than Kodaira $I_0^*$ (which is $4$ D7s on an O7, from a type IIb
perspective) or are $I_0^*$, and then furthermore that there exists a limit
in moduli space in which all of these enhance to $I_n^*$. This
is a very strong condition to satisfy, and we find it extremely
implausible that it is satisfied very often, an expectation
that is buttressed by our results. It would be interesting
to study this in the future to determine whether there is
a no-go for this to occur once an NHC exists.

More broadly, non-Higgsable clusters clearly play a role in obstructing weak coupling limits. However, the precise details
of the microphysics that does so is not
known, even though the mathematics is clear. It would be interesting to further understand the physics of the obstruction, including the interpretation of the exceptional divisors that arise from crepant resolutions of ADE singularities. Uncovering this mechanism is likely an important step in fully understanding strong coupling in F-theory.

\section*{Acknowledgements} We thank Massimo Bianchi, I\~naki Garc\'ia Etxebarria, Ben Heidenreich, Liam McAllister, and John Stout for useful discussions.
J.H. is supported by
NSF Grant PHY-1620526. B.S. is supported by NSF RTG Grant DMS-1645877.

\bibliographystyle{JHEP}
\bibliography{refs}

\providecommand{\href}[2]{#2}\begingroup\raggedright\begin{thebibliography}{10}

\bibitem{Bousso:2000xa}
R.~Bousso and J.~Polchinski, \emph{{Quantization of four form fluxes and
  dynamical neutralization of the cosmological constant}},
  \href{http://dx.doi.org/10.1088/1126-6708/2000/06/006}{\emph{JHEP} {\bf 06}
  (2000) 006}, [\href{http://arxiv.org/abs/hep-th/0004134}{{\tt
  hep-th/0004134}}].

\bibitem{Ashok:2003gk}
S.~Ashok and M.~R. Douglas, \emph{{Counting flux vacua}},
  \href{http://dx.doi.org/10.1088/1126-6708/2004/01/060}{\emph{JHEP} {\bf 01}
  (2004) 060}, [\href{http://arxiv.org/abs/hep-th/0307049}{{\tt
  hep-th/0307049}}].

\bibitem{Denef:2004ze}
F.~Denef and M.~R. Douglas, \emph{{Distributions of flux vacua}},
  \href{http://dx.doi.org/10.1088/1126-6708/2004/05/072}{\emph{JHEP} {\bf 05}
  (2004) 072}, [\href{http://arxiv.org/abs/hep-th/0404116}{{\tt
  hep-th/0404116}}].

\bibitem{MayorgaPena:2017eda}
D.~K. Mayorga~Pena and R.~Valandro, \emph{{Weak coupling limit of F-theory
  models with MSSM spectrum and massless U(1)'s}},
  \href{http://arxiv.org/abs/1708.09452}{{\tt 1708.09452}}.

\bibitem{Vafa:1996xn}
C.~Vafa, \emph{{Evidence for F theory}},
  \href{http://dx.doi.org/10.1016/0550-3213(96)00172-1}{\emph{Nucl. Phys.} {\bf
  B469} (1996) 403--418}, [\href{http://arxiv.org/abs/hep-th/9602022}{{\tt
  hep-th/9602022}}].

\bibitem{Morrison:1996pp}
D.~R. Morrison and C.~Vafa, \emph{{Compactifications of F theory on Calabi-Yau
  threefolds. 2.}},
  \href{http://dx.doi.org/10.1016/0550-3213(96)00369-0}{\emph{Nucl. Phys.} {\bf
  B476} (1996) 437--469}, [\href{http://arxiv.org/abs/hep-th/9603161}{{\tt
  hep-th/9603161}}].

\bibitem{Sen:1996vd}
A.~Sen, \emph{{F theory and orientifolds}},
  \href{http://dx.doi.org/10.1016/0550-3213(96)00347-1}{\emph{Nucl. Phys.} {\bf
  B475} (1996) 562--578}, [\href{http://arxiv.org/abs/hep-th/9605150}{{\tt
  hep-th/9605150}}].

\bibitem{Sen:1997bp}
A.~Sen, \emph{{Orientifold limit of F theory vacua}},
  \href{http://dx.doi.org/10.1016/S0920-5632(98)00143-1}{\emph{Nucl. Phys.
  Proc. Suppl.} {\bf 68} (1998) 92--98},
  [\href{http://arxiv.org/abs/hep-th/9709159}{{\tt hep-th/9709159}}].

\bibitem{Morrison:2012np}
D.~R. Morrison and W.~Taylor, \emph{{Classifying bases for 6D F-theory
  models}}, \href{http://dx.doi.org/10.2478/s11534-012-0065-4}{\emph{Central
  Eur. J. Phys.} {\bf 10} (2012) 1072--1088},
  [\href{http://arxiv.org/abs/1201.1943}{{\tt 1201.1943}}].

\bibitem{Morrison:2012js}
D.~R. Morrison and W.~Taylor, \emph{{Toric bases for 6D F-theory models}},
  \href{http://dx.doi.org/10.1002/prop.201200086}{\emph{Fortsch. Phys.} {\bf
  60} (2012) 1187--1216}, [\href{http://arxiv.org/abs/1204.0283}{{\tt
  1204.0283}}].

\bibitem{Grassi:2014zxa}
A.~Grassi, J.~Halverson, J.~Shaneson and W.~Taylor, \emph{{Non-Higgsable QCD
  and the Standard Model Spectrum in F-theory}},
  \href{http://dx.doi.org/10.1007/JHEP01(2015)086}{\emph{JHEP} {\bf 01} (2015)
  086}, [\href{http://arxiv.org/abs/1409.8295}{{\tt 1409.8295}}].

\bibitem{Halverson:2015jua}
J.~Halverson and W.~Taylor, \emph{{$ {\mathrm{\mathbb{P}}}^1 $-bundle bases and
  the prevalence of non-Higgsable structure in 4D F-theory models}},
  \href{http://dx.doi.org/10.1007/JHEP09(2015)086}{\emph{JHEP} {\bf 09} (2015)
  086}, [\href{http://arxiv.org/abs/1506.03204}{{\tt 1506.03204}}].

\bibitem{Taylor:2015ppa}
W.~Taylor and Y.-N. Wang, \emph{{A Monte Carlo exploration of threefold base
  geometries for 4d F-theory vacua}},
  \href{http://dx.doi.org/10.1007/JHEP01(2016)137}{\emph{JHEP} {\bf 01} (2016)
  137}, [\href{http://arxiv.org/abs/1510.04978}{{\tt 1510.04978}}].

\bibitem{Halverson:2017ffz}
J.~Halverson, C.~Long and B.~Sung, \emph{{On Algorithmic Universality in
  F-theory Compactifications}},  \href{http://arxiv.org/abs/1706.02299}{{\tt
  1706.02299}}.

\bibitem{Halverson:2016vwx}
J.~Halverson, \emph{{Strong Coupling in F-theory and Geometrically
  Non-Higgsable Seven-branes}},
  \href{http://dx.doi.org/10.1016/j.nuclphysb.2017.02.014}{\emph{Nucl. Phys.}
  {\bf B919} (2017) 267--296}, [\href{http://arxiv.org/abs/1603.01639}{{\tt
  1603.01639}}].

\bibitem{Nakayama}
N.~Nakayama, \emph{On {W}eierstrass models},  in \emph{Algebraic geometry and
  commutative algebra, {V}ol.\ {II}}, pp.~405--431.
\newblock Kinokuniya, Tokyo, 1988.

\bibitem{Taylor:2012dr}
W.~Taylor, \emph{{On the Hodge structure of elliptically fibered Calabi-Yau
  threefolds}}, \href{http://dx.doi.org/10.1007/JHEP08(2012)032}{\emph{JHEP}
  {\bf 08} (2012) 032}, [\href{http://arxiv.org/abs/1205.0952}{{\tt
  1205.0952}}].

\bibitem{Morrison:2014era}
D.~R. Morrison and W.~Taylor, \emph{{Sections, multisections, and U(1) fields
  in F-theory}},  \href{http://arxiv.org/abs/1404.1527}{{\tt 1404.1527}}.

\bibitem{Martini:2014iza}
G.~Martini and W.~Taylor, \emph{{6D F-theory models and elliptically fibered
  Calabi-Yau threefolds over semi-toric base surfaces}},
  \href{http://dx.doi.org/10.1007/JHEP06(2015)061}{\emph{JHEP} {\bf 06} (2015)
  061}, [\href{http://arxiv.org/abs/1404.6300}{{\tt 1404.6300}}].

\bibitem{Johnson:2014xpa}
S.~B. Johnson and W.~Taylor, \emph{{Calabi-Yau threefolds with large
  $h^{2,1}$}}, \href{http://dx.doi.org/10.1007/JHEP10(2014)023}{\emph{JHEP}
  {\bf 10} (2014) 23}, [\href{http://arxiv.org/abs/1406.0514}{{\tt
  1406.0514}}].

\bibitem{Taylor:2015isa}
W.~Taylor and Y.-N. Wang, \emph{{Non-toric bases for elliptic Calabi-Yau
  threefolds and 6D F-theory vacua}},
  \href{http://arxiv.org/abs/1504.07689}{{\tt 1504.07689}}.

\bibitem{Braun:2014xka}
A.~P. Braun and T.~Watari, \emph{{The Vertical, the Horizontal and the Rest:
  anatomy of the middle cohomology of Calabi-Yau fourfolds and F-theory
  applications}}, \href{http://dx.doi.org/10.1007/JHEP01(2015)047}{\emph{JHEP}
  {\bf 01} (2015) 047}, [\href{http://arxiv.org/abs/1408.6167}{{\tt
  1408.6167}}].

\bibitem{Watari:2015ysa}
T.~Watari, \emph{{Statistics of F-theory flux vacua for particle physics}},
  \href{http://dx.doi.org/10.1007/JHEP11(2015)065}{\emph{JHEP} {\bf 11} (2015)
  065}, [\href{http://arxiv.org/abs/1506.08433}{{\tt 1506.08433}}].

\bibitem{Halverson:2016tve}
J.~Halverson and J.~Tian, \emph{{Cost of seven-brane gauge symmetry in a
  quadrillion F-theory compactifications}},
  \href{http://dx.doi.org/10.1103/PhysRevD.95.026005}{\emph{Phys. Rev.} {\bf
  D95} (2017) 026005}, [\href{http://arxiv.org/abs/1610.08864}{{\tt
  1610.08864}}].

\bibitem{Morrison:2014lca}
D.~R. Morrison and W.~Taylor, \emph{{Non-Higgsable clusters for 4D F-theory
  models}}, \href{http://dx.doi.org/10.1007/JHEP05(2015)080}{\emph{JHEP} {\bf
  05} (2015) 080}, [\href{http://arxiv.org/abs/1412.6112}{{\tt 1412.6112}}].

\bibitem{Taylor:2015xtz}
W.~Taylor and Y.-N. Wang, \emph{{The F-theory geometry with most flux vacua}},
  \href{http://dx.doi.org/10.1007/JHEP12(2015)164}{\emph{JHEP} {\bf 12} (2015)
  164}, [\href{http://arxiv.org/abs/1511.03209}{{\tt 1511.03209}}].

\bibitem{Carifio:2017bov}
J.~Carifio, J.~Halverson, D.~Krioukov and B.~D. Nelson, \emph{{Machine Learning
  in the String Landscape}},
  \href{http://dx.doi.org/10.1007/JHEP09(2017)157}{\emph{JHEP} {\bf 09} (2017)
  157}, [\href{http://arxiv.org/abs/1707.00655}{{\tt 1707.00655}}].

\bibitem{Blumenhagen:2009up}
R.~Blumenhagen, T.~W. Grimm, B.~Jurke and T.~Weigand, \emph{{F-theory uplifts
  and GUTs}},
  \href{http://dx.doi.org/10.1088/1126-6708/2009/09/053}{\emph{JHEP} {\bf 09}
  (2009) 053}, [\href{http://arxiv.org/abs/0906.0013}{{\tt 0906.0013}}].

\bibitem{Grassi:2011hq}
A.~Grassi and D.~R. Morrison, \emph{{Anomalies and the Euler characteristic of
  elliptic Calabi-Yau threefolds}},
  \href{http://dx.doi.org/10.4310/CNTP.2012.v6.n1.a2}{\emph{Commun. Num. Theor.
  Phys.} {\bf 6} (2012) 51--127}, [\href{http://arxiv.org/abs/1109.0042}{{\tt
  1109.0042}}].

\bibitem{ROAN1996489}
S.~shyr Roan, \emph{Minimal resolutions of gorenstein orbifolds in dimension
  three},
  \href{http://dx.doi.org/http://dx.doi.org/10.1016/0040-9383(95)00018-6}{\emph{Topology}
  {\bf 35} (1996) 489 -- 508}.

\bibitem{roan1989}
S.-S. Roan, \emph{On the generalization of kummer surfaces},
  \href{http://dx.doi.org/10.4310/jdg/1214443600}{\emph{J. Differential Geom.}
  {\bf 30} (1989) 523--537}.

\bibitem{Joyce2003}
D.~Joyce, \emph{Riemannian Holonomy Groups and Calibrated Geometry}, pp.~1--68.
\newblock Springer Berlin Heidelberg, Berlin, Heidelberg, 2003.

\bibitem{Degeratu_flopsof}
A.~Degeratu, \emph{Flops of crepant resolutions}, {\emph{Turkish J. Math}
  23--40}.

\bibitem{2015arXiv150201040F}
L.~Facchini, V.~Gonz\'alez-Alonso and M.~Laso\'n, \emph{{Cox rings of du Val
  singularities}}, {\emph{ArXiv e-prints} (feb, 2015) },
  [\href{http://arxiv.org/abs/1502.01040}{{\tt 1502.01040}}].

\end{thebibliography}\endgroup

\end{document}